# PivCo-Huffman

**Marcin Żukowski**

*v.1.0, 2026-06-04*[1]

## Abstract

Huffman encoding has been an *enduring* technique for 70+ years, ubiquitous in compression algorithms since its invention. In this paper we propose a new approach to Huffman coding, based on a data structure from *wavelet trees*. The resulting *pivot-coded Huffman* (**PivCo-Huffman**) enables high-performance SIMD-friendly encoding and decoding operations. In our tests PivCo-Huffman consistently outperforms state-of-the-art Huffman codecs in decoding throughput. Additionally, we show how ANS-coding can be *selectively* applied to *skewed* nodes in this structure, yielding compression ratios approaching those of ANS-based codecs while preserving very high decompression speeds.

## 1. Introduction

Huffman encoding [1] is one of the most important algorithms in the area of compression. Moffat and Turpin nicely put it in [2] - it is very *enduring*: despite the introduction of better-compressing encodings (e.g. [3] or [4]), 70+ years on, it's still ubiquitous.

Note: formally, most modern systems don't necessarily use the *exact* encoding proposed in [1], but rather "canonical" coding from [5].

### 1.1. Classical Huffman tree

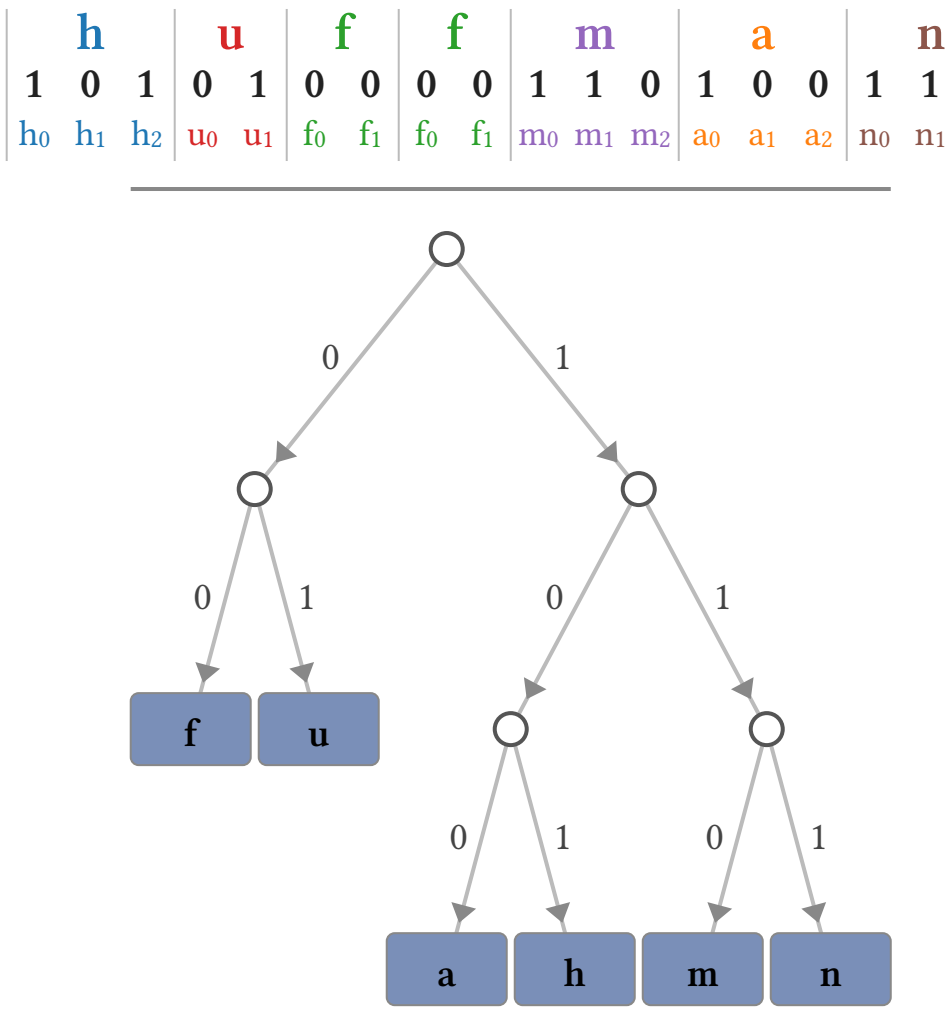

Figure 1: Classical Huffman tree for the word "huffman"

```
node = root

while not is_leaf(node)
  if read_bit() == 1:
    node = node->right
  else:
    node = node->left
return node.symbol
```

Listing 1: Naive Huffman decoding for one symbol

In classical Huffman coding, each symbol is encoded using a code (a sequence of bits), with more frequent symbols getting shorter codes. Figure 1 shows a Huffman tree for the word "huffman", and Listing 1 shows a naive decoding algorithm for decoding one symbol. For example, to decode symbol **"h"**, we traverse the tree using bits **"1 0 1"** to get to the proper leaf node representing that symbol.

[1] An interactive HTML version of this paper is available at https://marcinzukowski.github.io/pivco-huffman/paper-1.0//ph.html. Comments are welcome there.

## 1.2. Modern Huffman solutions

Implementation from Listing 1 is not very performant, as it uses a lot of operations and is not friendly for modern CPUs. Instead, modern Huffman decoding implementations use a *decoding table*, which allows decoding an entire symbol without traversing its code bit by bit. The size of supported code lengths is typically constrained, e.g. to *L=11* bits. Then a table of size *2^L* is created, allowing the following implementation:

```
code_bits = peek_bits(L);
emit_symbol(decoding_table[code_bits].symbol);
skip_bits(decoding_table[code_bits].numBits);
```

Such code can be further accelerated by using multiple cursors ([6], [7]), or by building a table that decodes two symbols in one iteration instead of one.

We measured various Huffman decoding implementations, and the most performant solutions we found were:

- **Huff0** - part of the open-source FSE ([8]) library, which is also a building block of the popular zstd compression library ([9]). Implemented in pure C, permissive license.
- **Oodle Huffman** - Huffman decoder from Oodle ([10]) - a proprietary compression library by RAD Game Tools. Implemented in C with a lot of assembly optimizations. Oodle requires a license for most uses.

Here are the measured bandwidths on two example datasets on two hosts (see Appendix C for more info):

| Dataset | Host | Huff0 | | Oodle Huffman | |
|---|---|---|---|---|---|
| | | enc MB/s | dec MB/s | enc MB/s | dec MB/s |
| proba80 | M4 | 1334 | 2929 | 1740 | 3343 |
| | c8i | 1147 | 1936 | 1295 | 2131 |
| prose_pride | M4 | 1722 | 2547 | 2325 | 3158 |
| | c8i | 1267 | 1758 | 1364 | 2120 |

These are impressive results. Still, in this paper we investigate if the performance could be further improved by using a completely different approach.

## 1.3. Motivating Example: Hash Join in Databases

Hash table lookup is one of the most performance-intensive operations in many systems, including databases. Below, we can see the pseudocode of a simple linear-hashing lookup:

```
hash = compute_hash(key)
pos = hash_table_first(hash)
while not hash_table_empty(pos)
  val = hash_table_value(pos)
  if val == key:
    return true
  pos = hash_table_next(pos)
return false
```

Just like Huffman decoding from Listing 1, this problem can be seen as a state-machine traversal. In both cases, data dependencies in the loop and unpredictable branching prevent the CPU from achieving high performance. Hash join additionally performs an expensive memory lookup causing additional stalls.

[11] (Section 5.3.3.2) proposed an alternative hash table lookup approach based on the idea of going through each node in the state machine not for one, but for a *vector* of records, presented in this pseudocode:

```
misses = []                          // miss input positions
hits = []                            // hits input positions
hash = compute_hash(keys)            // all input hash values
active = hash_table_first(hash)      // input positions we're still looking up
while not active.empty():            // if we still have work to do
  // move empty slots to misses, reduce active
  hash_table_split_empty(&active, &misses)
  // get all values from the hash table for active indices
  vals = hash_table_vals(active)
  // compute comparisons
  comp_results = compare(vals, keys, active)
  // split into hits if equal, active if not - those need more work
  split_on_equality(comp_results, &active, &hits)
  // get all the next positions for all still active records
  active = hash_table_next(active)
// misses have all miss positions, hits have all hit positions
```

This approach, while more complex and seemingly labor-intensive (definitely issues more CPU instructions), in each phase exposes to CPUs a lot of simple, independent operations, avoids any data or control dependencies, and allows overlapping memory accesses. As a result, it achieves a significant performance benefit (even >10x) over the *scalar* approach.

## 2. Pivoting Huffman

Following the example from Section 1.3, it should be possible to create a Huffman decoder using similar principles. However, for each node in the Huffman tree to have access to relevant bits from the stream, a different data layout is needed. We found a great solution for this in the *wavelet tree* structure [12].

Figure 2 shows an alternative representation of the encoded word “huffman”. Instead of a code-after-code stream, we divide all the stream bits by their (possibly empty) prefix. Figure 3 shows how this layout maps onto the Huffman tree when decoding data. Each node receives all the bits of the codes that pass through it, and navigates these codes to its children, where another bitmap is used for the next step.

Note that while logically this representation contains the same information as standard Huffman coding, it typically stores bitmaps byte-aligned. This might lead to a *marginally worse* compression ratio due to byte-padding. However, for non-trivial datasets this overhead is acceptable if this approach provides other benefits.

This data representation and tree traversal are the basis for ***pivot-coded Huffman*** **(PivCo-Huffman)**, presented in this paper. In this section we present the *initial* implementation of this approach, which is actually *not* used in the final solution. Still, since it is a more *natural* approach, we describe it first, and use it to introduce a set of optimizations and implementation techniques. In Section 3, we will propose the final, more performant solution.

Figure 2: Example of pivoting a Huffman-encoded string. All bits sharing the same prefix (noted in quotes) are grouped together as a bitmap. Color-coded letters with subscript denote which-letter which-bit combinations. * marks a code-terminal bit.

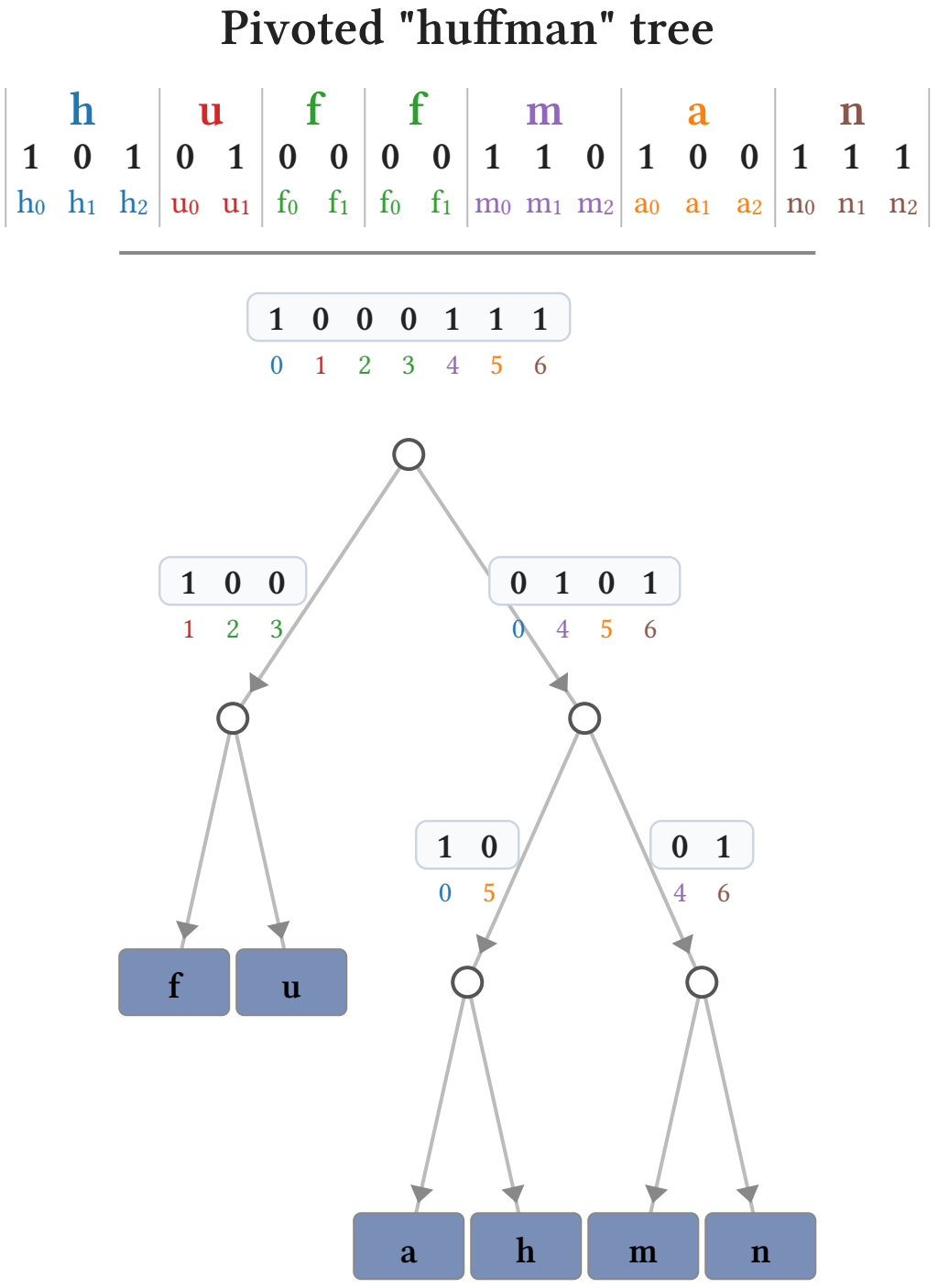

Figure 3: Example of a Huffman tree with "pivoted" data traversing it, reusing the previous example. Each node produces a list of indices for its symbols.

As mentioned, the tree representation used is equivalent to *wavelet trees*, specifically *Huffman-shaped wavelet trees* (e.g. [13]). However, we treat PivCo-Huffman as a related, but separate solution - see Section 7.2 for more discussion.

## 2.1. Naive implementation

With a defined Huffman tree, and data stored in per-node bitmaps, we can traverse the tree top-down. Note, as we do it, we need to know which output elements we are decoding. For that, we carry an additional `indices` list (the root node does not need it). In our implementation we use 16-bit values for indices, as we decode data in small blocks (e.g. 8KB). With that, PivCo-Huffman tree traversal boils down to applying two operations:

**`partition(bitmap, indices) => (indices_left, indices_right)`** – applied for all internal nodes. Takes a list of indices (positions in the output stream), and divides it based on the bitmap into left and right indices for its subtrees. Note, a special `partition_root` version can be used in the root node, as its list of indices is the complete input.

The `partition` primitive can be expressed naively with:

```
for (i = 0; i < n; i++) {
  bit = get_bit(bitmap, i);
  if (bit) indices_right[n_right++] = indices[i];
  else     indices_left[n_left++] = indices[i];
}
```

**`scatter(output, indices, symbol)`** – fills all positions in the output stream with a given symbol.

```
for (i = 0; i < n; i++) {
  output[indices[i]] = symbol;
}
```

We measured the decoding performance of such a naive implementation on Apple M4 CPU, and, as expected, the performance is very sub-par.

| **Data** | **M4** (GB/s) | | | **c8i** (GB/s) | | |
|---|---|---|---|---|---|---|
| | naive | Huff0 | naive/Huff0 | naive | Huff0 | **naive/Huff0** |
| proba80 | 1.38 | 2.93 | 0.47 | 0.35 | 1.94 | **0.18** |
| prose_pride | 0.45 | 2.55 | 0.18 | 0.06 | 1.76 | **0.03** |

There are two main reasons for this:

- for each decoded symbol, we perform multiple operations: we run `partition` for each bit in the code, followed by a final `scatter` at each leaf (`len(code)+1` operations in total).
- the `partition` and `scatter` primitives, as written, are not efficient.

In the following two sections we will discuss how to address both problems.

## 2.2. Tree Optimizations

A naive Huffman tree discussed before suffers from a large number of operations per byte. In this section we demonstrate a number of techniques that can bring that number down significantly, using string `coconut-papaya` as a test case.

In Figure 4 we see our starting point - a basic tree with 2 kinds of primitives (marked in orange boxes), and 4.071 operations per output byte (weighted by symbol frequency). Table 1 translates symbols used in figures in this section to the actual compute primitives.

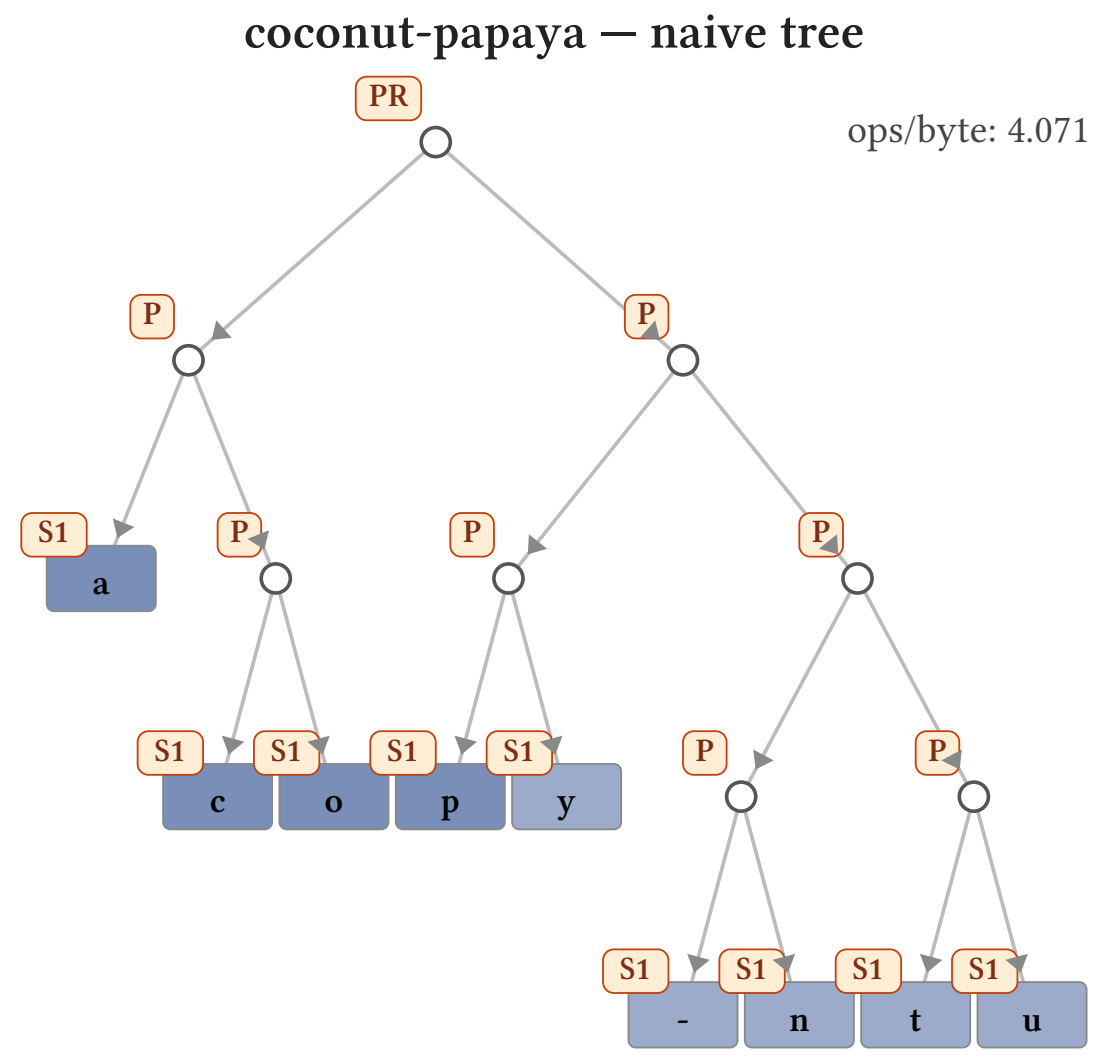

Figure 4: Decoding-strategy for a naive Huffman tree

| **Code** | **Explanation** |
|---|---|
| P | `partition` - split indices into left/right based on bitmap |
| PR | `partition_root` - like `partition` but for the root node |
| PH | `partition_half` - like `partition`, but produces only one output |
| C | *not a primitive* - marks the “constant”, top-frequency key |
| S1 | `scatter` - scatters a single symbol into output |
| S2 | `scatter_two` - scatters two symbols into output |
| SFD | `scatter_flat_D` - scatters 2^D symbols into output |

Table 1: Primitive symbols used in figures in this Section

### 2.2.1. Merging leaves

One simple approach of reducing the number of operations is to avoid the last-level `partition`, and simply fill all input indices with the symbol based on the bitmap.

This results in a **`scatter_two(output, indices, bitmap, symbol0, symbol1)`** primitive:

```
for (i = 0; i < n; i++) {
  output[indices[i]] = get_bit(bitmap, i) ? symbol1 : symbol0;
}
```

Figure 5 shows the benefit in reduced operations per decoded byte going from 4.071 to 3.286.

### 2.2.2. Frequent symbol optimization

One of the problems of our decoding primitives is writing into non-contiguous positions in the output, presenting challenges for modern CPUs and memory subsystems (see Section 2.4.2).

We can mitigate it by avoiding this completely for the *most frequent symbol*, by simply prefilling the entire output with `memset` before decoding. Then, that symbol never needs to be processed during the tree traversal. Note that `memset` is 1-2 orders of magnitude faster than our primitives, so that cost is negligible.

Figure 6 shows the tree with this optimization applied to symbol **`"a"`**. Note that as a result we introduce a new operation called `PH` - `partition_half`, similar to `partition`, but only producing one of the output index lists.

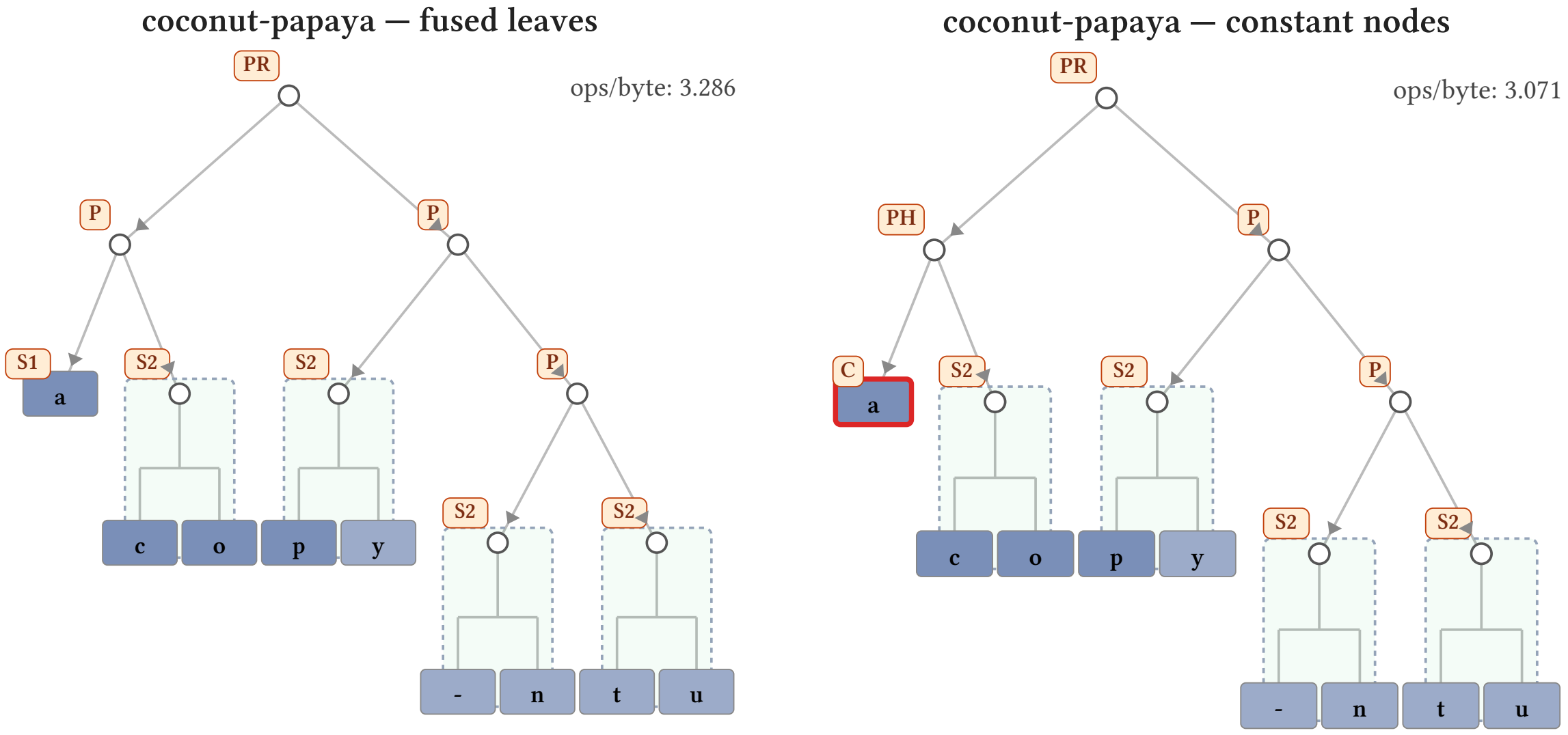

Figure 5: Merging `partition` and `scatter` into `scatter-two`

Figure 6: Pre-filling the most frequent "constant" symbol

### 2.2.3. Flat Subtrees

Huffman trees often contain subtrees where all the symbols share the same length. In our example, 4 right-most **`"- n t u"`** nodes form such a subtree.

We can decode such a subtree with a single operation **`scatter_flat_D(output, indices, bitmap, symbols)`**, where `D` represents the depth of the subtree.

Note that for this, the input `bitmap` is not *binary*, but *(2^D)-ary*, with bits packed contiguously. Also, note that `scatter_two` is a special case of this approach, with *D=1*.

```
bit_unpack(bitmap, D, code_indices);
for (i = 0; i < n; i++) {
  output[indices[i]] = code_to_symbols[code_indices[i]];
}
```

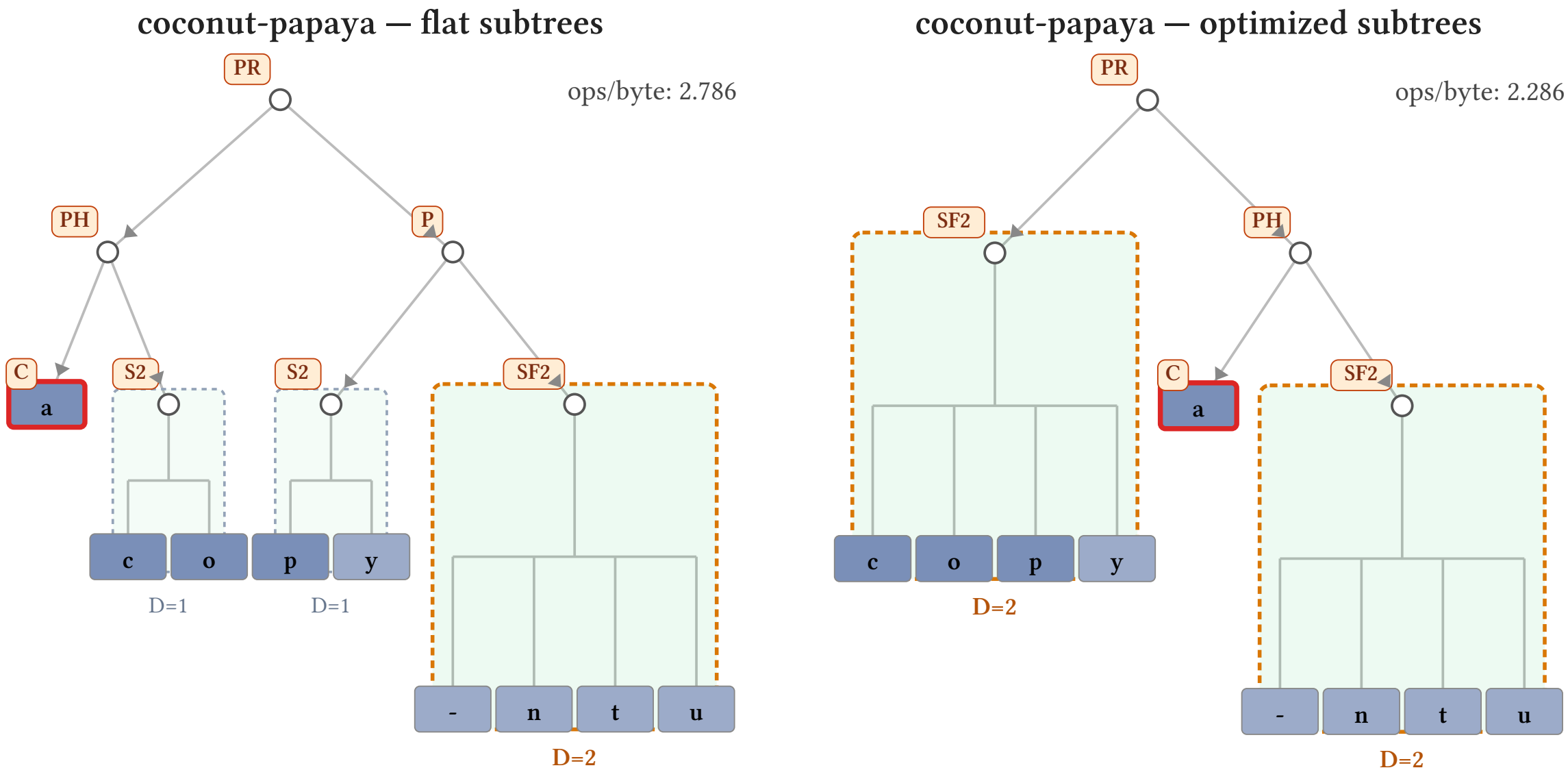

Figure 7: Detecting “flat” subtrees

Figure 8: Optimizing “flat” subtrees

#### 2.2.4. Non-Canonical Subtrees

Looking at Figure 7, we can see that while the `"- n t u"` symbols benefit from the “flat subtrees” strategy, we also have `"c o p y"` symbols, which share the same code lengths, but are not decoded together.

We can reorganize the canonical Huffman tree to make it more amenable to the “flat subtree” optimization by making sure that codes with the same length are grouped as much as possible. To achieve that, after determining code lengths, within each length-group, we combine the largest *power of two* number of nodes into a single node with a combined frequency. We repeat the process, with one length-group possibly creating multiple such nodes (of different depth).

The result is a new, (usually) non-canonical Huffman tree, with the exact same average code lengths, but a different shape. Figure 8 shows how applying this strategy allows the `"c o p y"` nodes to be processed together, further reducing ops/byte.

### 2.3. Impact of tree optimizations

| **Data** | **M4** (GB/s) | | | | **c8i** (GB/s) | | | |
|---|---|---|---|---|---|---|---|---|
| | naive | opt | Huff0 | opt/Huff0 | naive | **opt** | Huff0 | opt/Huff0 |
| proba80 | 1.38 | 1.73 | 2.93 | 0.59 | 0.35 | **0.84** | 1.94 | 0.43 |
| prose_pride | 0.45 | 0.6 | 2.55 | 0.24 | 0.06 | **0.08** | 1.76 | 0.04 |

Table 2: Combined impact of tree optimizations on performance

Table 2 shows that applying all optimizations above allows improving PivCo-Huffman performance by up to factor two. However, as we will demonstrate later, these optimizations are even more important with faster compute primitives.

### 2.4. Computing Primitives

Table 1 lists primitives used during decoding of PivCo-Huffman. Most of them can be expressed with a simple “scalar” loop, but since they work on multiple items, they also represent opportunities for SIMD optimization. And even with a scalar loop, coding methods can make a dramatic difference.

In this Section we’ll discuss how some of them are implemented using SIMD instructions from the ARM NEON family.

| Primitive | proba80 | | prose_pride | |
|---|---|---|---|---|
| | M4 | c8i | M4 | c8i |
| p_partition | 0.40 | 2.08 | 0.40 | 4.23 |
| p_half_right | 0.37 | 0.65 | 0.41 | 0.66 |
| s1_scatter | 0.26 | 0.21 | 0.28 | 0.30 |
| s2_scatter_both | 0.51 | 3.47 | 0.32 | 4.22 |
| sfx_scatter_flat | — | — | 0.39 | 0.93 |

Table 3: Performance of “naive” primitive implementations (ns/code)

#### 2.4.1. `partition`

`partition` is the most important operation during PivCo-Huffman tree traversal, as each code goes through it multiple times before ending up in one of the versions of `scatter`.

A naive implementation of this primitive was presented in Section 2.1. The critical performance aspect there is this fragment:

```
    if (bit) indices_right[n_right++] = indices[i];
    else     indices_left[n_left++] = indices[i];
}
```

This statement depends on the extracted bit value, and may be hard to predict for the branch predictor. If the distribution is skewed (like `proba80`), it makes the branch easier to guess. For more uniform data, the branch predictor cannot guess properly, as we can see in Table 3 for `prose_pride`. On `m4`, with its cheaper branch predictor misses, the difference is minimal, but on `c8i` it is significant.

One way to alleviate this is to replace branching with an unconditional assignment, here’s this approach shown inside the smaller `partition_half_right` primitive:

```
    // ... same setup
    right[n_right] = indices[i];
    n_right += b;
```

In Table 3 we see that this primitive doesn’t suffer from the branch misprediction on `prose_pride` as much as `partition`.

Partitioning performance can be further improved with SIMD. For example, here’s an ARM NEON implementation (just an 8-value kernel with a given 8-bit `mask` from the bitmap).

```
uint8x16_t data = vld1q_u8((const uint8_t *)indices);

/* Load shuffle patterns for right/left side - they are stored together */
const uint8_t *tab = compress_tab[mask];
```

```
uint8x16_t shuf_r = vld1q_u8(tab);       /* bytes 0-15: right */
uint8x16_t shuf_l = vld1q_u8(tab + 16);  /* bytes 16-31: left */

/* Save input indices in either right or left */
uint8x16_t right = vqtbl1q_u8(data, shuf_r);
uint8x16_t left  = vqtbl1q_u8(data, shuf_l);

/* Compute how many values went right */
int n_right = compress_popcnt[mask];

/* Store both results - always 16 bytes */
vst1q_u8((uint8_t *)right_out, right);
vst1q_u8((uint8_t *)left_out, left);
```

How it works:

- `compress_tab`, for each of the 256 possible `mask` values, stores two (for left and right) 16-byte arrays determining which bytes from the input should be written to a given output
- `vqtbl1q_8` operation creates *condensed* subsets of input for left/right. Note, these vectors might have zeros in their tail, as on average they are half-full.
- result vectors are always written as 16 bytes - while it might sound wasteful, it is simpler and faster.
- the next iteration needs to adjust both output pointers based on `n_right` (`n_left` is simply `8 - n_right`)

Note that this code is fully branch-free, and with just a few instructions we process 8 16-bit input indices. We can apply a similar strategy on other architectures. For example, on AVX-512 one can partition even 32 16-bit index values in a few steps like that. The results in Table 4 show how the performance of the `partition` primitive improved by a factor of **6x** on **m4** and **25-70x** (!) on **c8i**.

#### 2.4.2. `scatter` primitives

The other big part of tree traversal are `scatter` primitives. They come in a few forms:

- `S1` - scatter_one - puts a constant symbol at all input indices
- `S2` - scatter_two - puts one of two symbols based on the bitmap
- `SFD` - scatter_flat_D - puts one of the 2^D symbols in the indices based on a packed x-bit bitmap.

Each of these primitives can be decomposed into two stages:

1. Determine which symbol to write at a given index
2. Actually write the symbols

For `scatter_one`, part one is trivial. For `scatter_two`, it's a bit more interesting. One could of course do code similar to

```
output[indices[i]] = bitmap[i] ? symbol0 : symbol1
```

For an efficient SIMD implementation we prefer another approach, based on SIMD arithmetic and a precomputed delta between `symbol0` and `symbol1`.

```
// Performed once
uint8x8_t vsym0  = vdup_n_u8(sym0);          // vector of sym0
uint8x8_t vdelta = vdup_n_u8(sym0 ^ sym1);    // vector of sym0^sym1
static const uint8_t bit_pos_tab[8] = {1,2,4,8,16,32,64,128};
```

```
  uint8x8_t vbit_pos = vld1_u8(bit_pos_tab);    // vector used to "expand a
bitmap"

  // Performed for each bitmap byte
  // convert 8-bits into 8 00/FF bytes
  uint8x8_t bits = vtst_u8(vdup_n_u8(bitmap[i >> 3]), vbit_pos);
  // set 8 symbols to either sym0 or sym1 (=sym0^delta)
  uint8x8_t vals = veor_u8(vsym0, vand_u8(vdelta, bits));
  // write 8 symbols
  output[indices[i+0]] = vget_lane_u8(vals, 0);
  // ...
  output[indices[i+7]] = vget_lane_u8(vals, 7);
```

For `scatter_flat_D` we get a D-bit packed bitmap. The first step is to unpack it, using an optimized unpacking kernel (see Section 7.3). Then each unpacked value can be used to lookup an actual symbol to write from a table. On ARM this can be done with `vqtbl*` instructions. For example, here's an implementation for D=5:

```
  // Performed once.
  // Assumes c2s is an array mapping index 0..31 to a given symbol.
  // We load these values into two uint8x16 vectors.
  uint8x16x2_t c2s_vec;
  c2s_vec.val[0] = vld1q_u8(c2s);
  c2s_vec.val[1] = vld1q_u8(c2s + 16);

  // Performed for each group of 8 packed values
  uint8x8_t codes = flat_d5_unpack(bitmap + ((i * 5) >> 3));  // unpack
  uint8x8_t vals  = vqtbl2_u8(c2s_vec, codes);                 // lookup
  // ... write 8 symbols - identical as in scatter_two
```

The second problem is the actual writing of the symbols. It boils down to the following problem:

```
  output[indices[i+0]] = input_function(i+0)
  // ...
  output[indices[i+7]] = input_function(i+7)
```

Note that with indices being ordered, but not contiguous, this results in a lot of individual writes. All 3 versions of `scatter` use this approach. This tends to saturate the CPU's load/store units, and limits further performance improvements. The author does not know of an efficient solution to this problem on either x86 or ARM architectures. Only AVX-512 provides *scatter* instructions, but they do not seem applicable here, as they only work with 32- and 64-bit values.

Table 4 demonstrates the memory-writes problem. You can see even the seemingly trivial `simd_s1_scatter` taking 2-5x more time per element than `simd_partition`. We also see that SIMD optimizations only improved scatter performance by up to a factor of 2x compared to Table 3.

Note that per-dataset numbers vary due to different cardinalities. In particular, `proba80` has very few elements reaching `simd_s2_scatter_both`, causing a high per/element cost.

### 2.5. Results

Table 5 and Table 6 show how thanks to combining tree optimizations and high-performance SIMD primitives, this version of PivCo-Huffman enters the performance territory of Huff0. We also see how with faster primitives the impact of the optimized tree shape provides stronger and more consistent benefits compared to Table 2.

| Primitive | proba80 | | prose_pride | |
|---|---|---|---|---|
| | M4 | c8i | M4 | c8i |
| simd_partition | 0.063 | 0.081 | 0.064 | 0.064 |
| simd_partition_half_right | 0.049 | 0.036 | 0.048 | 0.035 |
| simd_s1_scatter | 0.134 | 0.165 | 0.152 | 0.217 |
| simd_s2_scatter_both | 0.219 | 1.324 | 0.163 | 0.235 |
| simd_sfx_scatter_flat | — | — | 0.161 | 0.220 |

Table 4: Performance of SIMD optimized primitive implementations (ns/code)

Notably, the performance really depends on dataset - in `proba80`, with its lower entropy / shorter codes, average number of operations per symbol is much smaller. This behavior is unique to PivCo-Huffman, and allows it to decidedly beat Huff0 on such distributions.

Still, the performance is not consistently impressive - this is mostly impacted by the bottleneck of writes in `scatter` primitives.

In the next Section we'll discuss a different approach to PivCo-Huffman that works around this problem.

| Dataset | Tree | M4 | | c8i | |
|---|---|---|---|---|---|
| | | scalar | simd | scalar | simd |
| proba80 | naive | 1.38 | 4.75 | 0.35 | 2.76 |
| | optimized | 1.73 | **9.66** | 0.84 | **8.85** |
| prose_pride | naive | 0.45 | 2.02 | 0.06 | 1.26 |
| | optimized | 0.6 | **2.79** | 0.08 | **1.9** |

Table 5: Impact of tree and primitive optimizations (GB/s)

| Dataset | Tree | M4 | | c8i | |
|---|---|---|---|---|---|
| | | scalar | simd | scalar | simd |
| proba80 | naive | 0.47 | 1.62 | 0.18 | 1.42 |
| | optimized | 0.59 | **3.3** | 0.43 | **4.57** |
| prose_pride | naive | 0.18 | 0.79 | 0.03 | 0.72 |
| | optimized | 0.24 | **1.1** | 0.04 | **1.08** |

Table 6: Impact of tree and primitive optimizations (ratio to Huff0, higher is better)

# 3. Going Bottom-Up

In Section 2 we processed the tree **top-down**, which is a very natural approach, directly translating to "textbook" Huffman decoding. Still, while achieving decent performance, it is heavily penalized by the high number of scattered writes.

When exploring solutions to this problem, we looked at an idea of first traversing the tree to identify index-symbol positions, then merging index positions back into a contiguous sequence, and then using that for a scatter-free writing of symbols. During the merging phase, we would need to carry per-position *symbols* together with the *indices* all the way back to the root. That particular idea was quickly dropped due to a high cost of merging added to already significant cost of tree traversal.

However, it led to another realization. We can use the idea of **bottom-up** merging without the first partitioning stage at all. This idea led to a new variant of PivCo-Huffman, which is the actual proposed solution. The process is as follows:

Every tree node produces the values for all output positions with symbols that traverse a given node - these were the indices in the top-down traversal. For leaves, all these values are constants. For non-leaf nodes, we can construct the output using children symbols, and the same bitmaps we used for **bitmap-based partitioning**, but now using **bitmap-based merging**. This process proceeds all the way to the top, resulting in the final sequence of codes equal to the complete expected output.[2]

In this approach, leaf nodes do not require any processing, as they just produce a constant value. This is different from the top-down approach, where we had to apply a `scatter` primitive. Additionally, inputs and output of each node are *dense*, alleviating the scatter problem of the top-down approach.

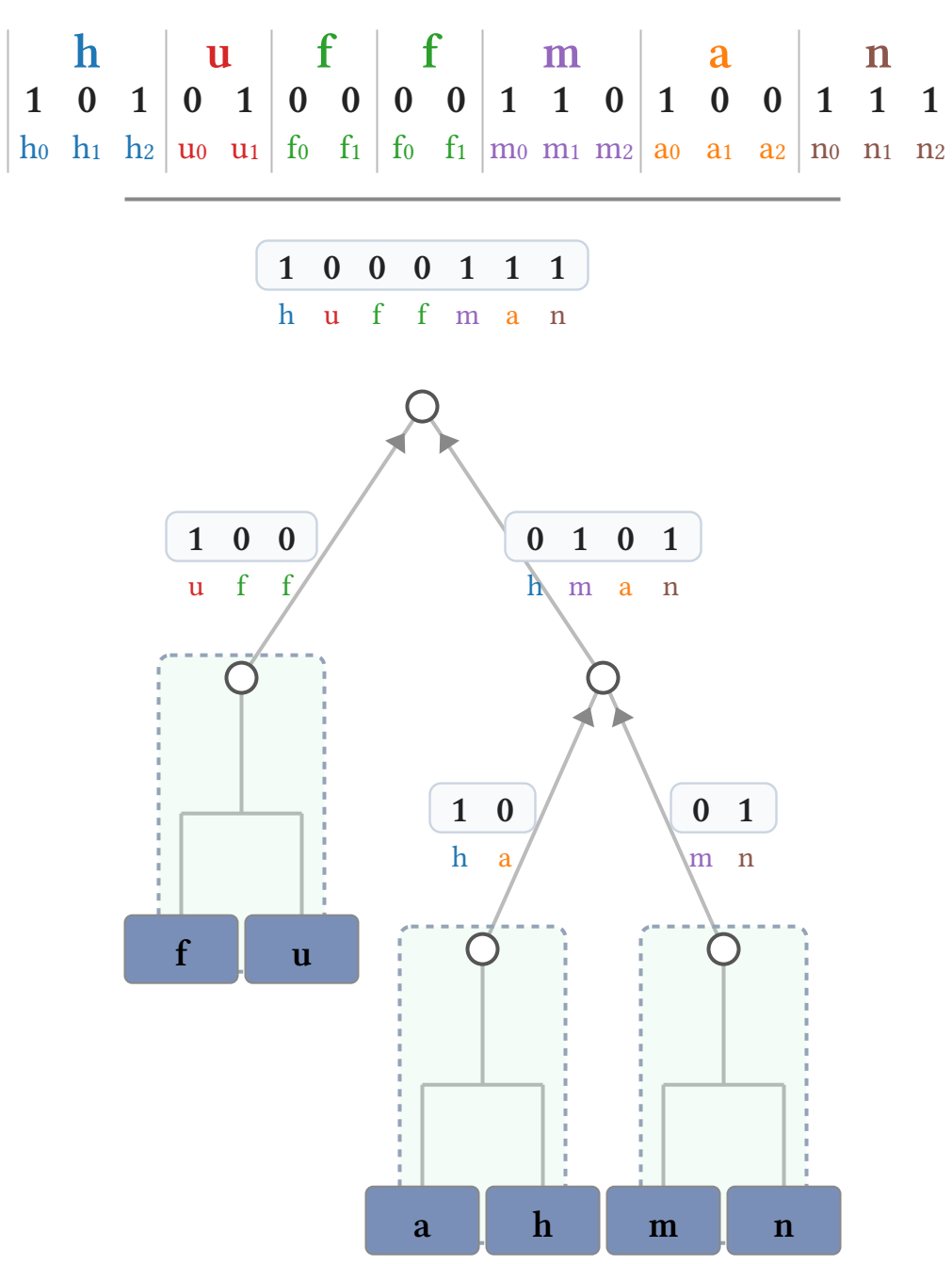

Figure 9: Bottom-up traversed "huffman" tree (*flat trees* off). See how each node produces a dense list of symbols.

This results in the approach presented in Figure 9. Note that this tree is symmetrical to Figure 3, with just data traveling in the opposite direction, and different data flowing with the Huffman-code bitmaps (symbols vs indices).

### 3.1. Bottom-up tree optimizations

With a different processing model, let us see how tree optimizations from Section 2.2 apply to the bottom-up approach:

- *fused-leaves* - not applicable, as the leaves just produce constant values
- *frequent-symbol* - not applicable, as the final merge in the root produces a dense full-output sequence
- *flat-subtrees* - directly applicable, reduces the tree size
- *non-canonical flat-subtrees* - directly applicable, further reduces the tree size

[2] Note that a similar symmetry of *partitioning* vs *merging* can be found in other places, e.g. sorting or joins in databases.

| Dataset | H | L | naive | | flat | | flat-opt | |
|---|---|---|---|---|---|---|---|---|
| | | | nodes | ops/B | nodes | ops/B | nodes | ops/B |
| proba80 | 0.905 | 1.251 | 5 | 1.251 | 5 | 1.251 | 5 | 1.251 |
| english | 4.226 | 4.248 | 29 | 4.248 | 21 | 3.700 | 15 | 3.012 |
| html_wiki | 5.472 | 5.629 | 196 | 5.629 | 34 | 4.407 | 30 | 4.036 |
| prose_pride | 4.531 | 4.596 | 95 | 4.596 | 33 | 3.896 | 31 | 3.865 |
| image_jpeg | 7.890 | 7.920 | 255 | 7.920 | 35 | 4.443 | 31 | 3.574 |
| json_api | 5.199 | 5.240 | 97 | 5.240 | 39 | 4.564 | 31 | 4.130 |
| dna_fasta | 2.081 | 2.266 | 37 | 2.266 | 15 | 2.265 | 13 | 2.265 |
| chinese_text | 5.795 | 5.951 | 149 | 5.951 | 33 | 4.662 | 25 | 3.968 |
| calgary_pic | 1.189 | 1.676 | 158 | 1.676 | 20 | 1.489 | 18 | 1.475 |

Table 7: The impact of tree optimizations on the decoding cost. **L** - the average (weighted) Huffman code symbol.

Table 7 shows the impact of flat-subtrees and fully-optimized non-canonical trees. The significant reduction in *ops/B* results in a better encoding and decoding performance. Additionally, the reduction of the number of nodes helps tree construction time and tree traversal overheads. See also Table 10 for the actual decoding performance.

### 3.2. Bottom-up tree operations

Bottom-up processing uses two families of *merge* operations. First are binary `merge_X_Y` primitives, where both `X` and `Y` can be `vec` (a vector of symbols) or `cst` (a constant symbol). The second family consists of N-ary `merge_flat_D` primitives, specialized for `D` values. Figure 10 shows how they are used to build the tree, and Table 8 discusses how these primitives correspond to the top-down primitives.

A naive implementation of e.g. `merge_vec_vec` would be directly symmetrical to `partition` from Section 2.1:

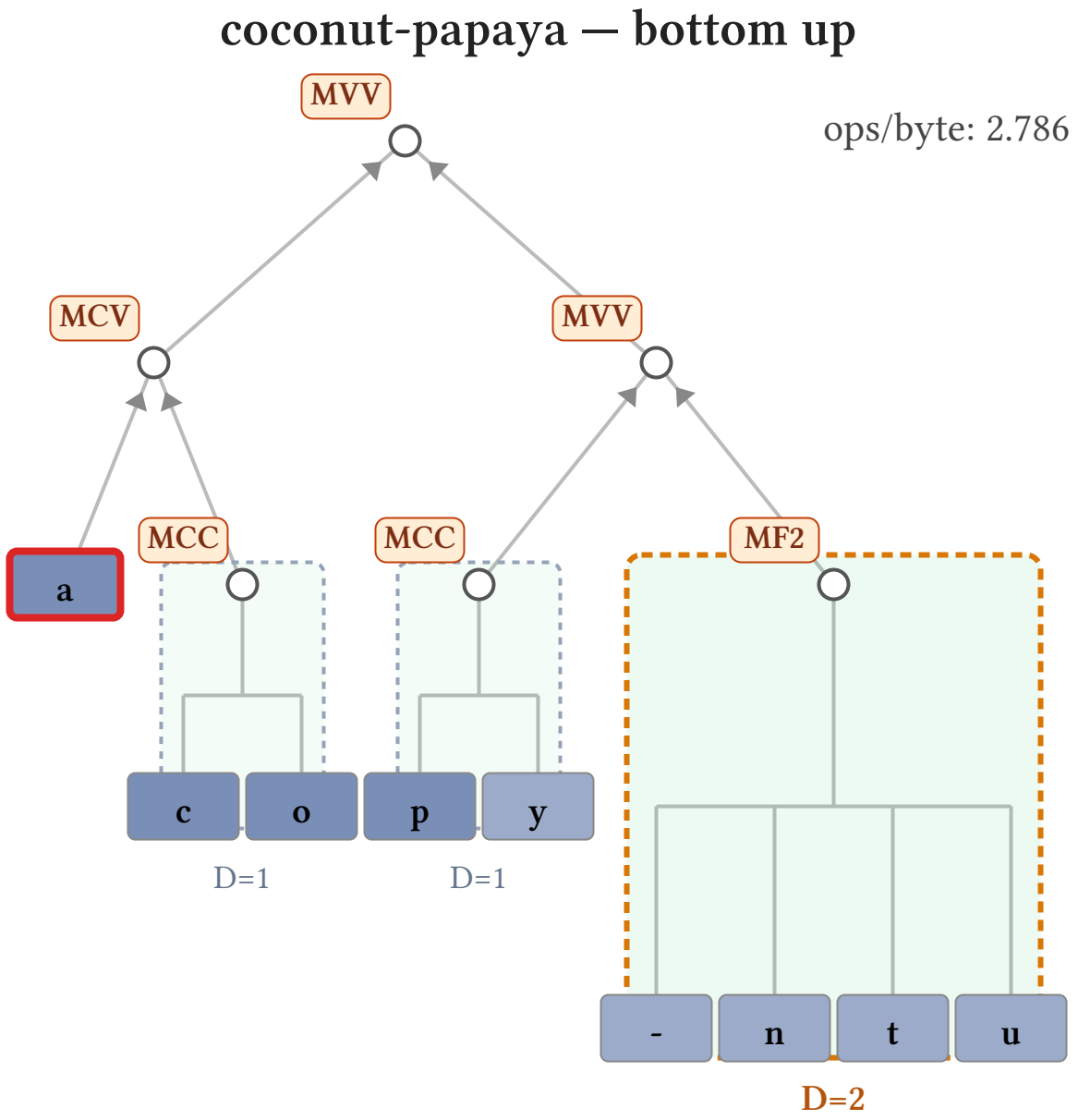

Figure 10: Bottom-up tree operations (*optimized flat trees* off)

| Top-down equivalent | Bottom-up operation | Explanation |
|---|---|---|
| `P` | `MVV` | `merge_vec_vec` is symmetrical to two-sided partition |
| `PR` | `MVV` | `merge_vec_vec` for the root node is identical to other cases |
| `PH` | `MCV/MVC` | `merge_cst_vec/merge_vec_cst` - *merge* variant where one input is constant |
| `S2` | `MCC` | `merge_cst_cst` - merges two constant symbols into output |
| `SFD` | `MFD` | `merge_flat_D` - merges 2^D constant symbols into output |
| `C` | – | Note that in bottom-up multiple leaves can be "constant" |
| `S1` | – | No operation needed for leaves when going bottom up |

Table 8: Primitives used in bottom-up processing and their top-down equivalents

```
for (i = 0; i < n; i++) {
  bit = get_bit(bitmap, i);
  if (bit) output[i] = symbols_right[n_right++]
  else     output[i] = symbols_left[n_left++]
```

Naturally, we implement this logic with SIMD, using the following code on ARM NEON, for 8 entries

```
uint8_t mask  = bitmap[i >> 3];
// Load eight left and right symbols
uint8x8_t  lsyms = vld1_u8(left + n_left);
uint8x8_t  rsyms = vld1_u8(right + n_right);
// Combine them into a single vector
uint8x16_t both = vcombine_u8(lsyms, rsyms);
// Load the precomputed shuffle vector for this mask
uint8x8_t  shuf = vld1_u8(expand_tab[mask]);
// Gather values from either left or right input based on mask
uint8x8_t  o    = vqtbl1_u8(both, shuf);
// Save 8 bytes, always
vst1_u8(out + i, o);
// Update n_left and n_right for the next iteration
int nr = expand_popcnt[mask];
n_right += nr;
n_left += (8 - nr);
```

The code for `merge_cst_vec` is identical, except we use a precomputed (outside the hot loop) vector of constant values, e.g.:

```
uint8x8_t  lsyms = vdup_n_u8(left_sym);
```

| primitive | | | | M4 | | | | c8i | | | |
|---|---|---|---|---|---|---|---|---|---|---|---|
| **name** | in_b | out_b | lut_b | **ns/el** | in_bw | out_bw | lut_bw | **ns/el** | in_bw | out_bw | lut_bw |
| **merge_vec_vec** | 17 | 8 | 9 | **0.046** | 46.4 | 21.8 | 24.5 | **0.026** | 82.6 | 38.9 | 43.7 |
| **merge_cst_cst** | 1 | 8 | 0 | **0.019** | 6.5 | 52.2 | 0.0 | **0.010** | 12.2 | 97.7 | 0.0 |
| **merge_cst_vec** | 9 | 8 | 9 | **0.040** | 27.9 | 24.8 | 27.9 | **0.017** | 65.2 | 58.0 | 65.2 |
| **merge_flat_d2** | 2 | 8 | 8 | **0.019** | 13.2 | 52.9 | 52.9 | **0.031** | 8.0 | 32.0 | 32.0 |
| **merge_flat_d3** | 3 | 8 | 8 | **0.042** | 8.9 | 23.8 | 23.8 | **0.030** | 12.6 | 33.6 | 33.6 |
| **merge_flat_d4** | 4 | 8 | 8 | **0.019** | 26.2 | 52.3 | 52.3 | **0.031** | 16.1 | 32.2 | 32.2 |
| **merge_flat_d5** | 5 | 8 | 8 | **0.039** | 16.2 | 25.9 | 25.9 | **0.036** | 17.2 | 27.5 | 27.5 |
| **merge_flat_d6** | 6 | 8 | 8 | **0.047** | 16.1 | 21.4 | 21.4 | **0.048** | 15.5 | 20.7 | 20.7 |
| **merge_flat_d7** | 7 | 8 | 8 | **0.073** | 12.0 | 13.8 | 13.8 | **0.064** | 13.6 | 15.5 | 15.5 |
| **merge_flat_d8** | 8 | 8 | 8 | **0.064** | 15.5 | 15.5 | 15.5 | **0.019** | 51.8 | 51.8 | 51.8 |

Table 9: Bottom-up primitive performance on M4 and c8i. **in_b / out_b / lut_b** - input /output / lookup table **bits** used per element. **in_bw / out_bw / lut_bw** - respective memory bandwidths achieved in **GB/s**.

`merge_vec_cst` is symmetrical. Other *merge* primitives do not have symbols as an input, but rather, logically, a bit-packed index into a *code-to-symbol* table. As such, the process for all these primitives consists of unpacking the packed-values, and then performing such a lookup. We found this solution to be the fastest even for `merge_cst_cst`, which can be seen as `merge_flat_D` with `D=1`.

For the lookup, on M4 we use the family of `vqtbl*` operations for D=1..6, chained with `vqtbx*` operations for D=7..8. Here's an example for D=4 (16 symbols):

```
// Before loop - load the code-to-symbol mapping into a vector
uint8x16_t c2s_vec = vld1q_u8(c2s);

// In a loop, for 16 (!) elements
// Unpack 16 nibbles (8 bytes) into 16 code-index bytes
uint8x16_t codes = flat_d4_unpack(bitmap + (i / 2));
// Fetch the symbols we need
uint8x16_t syms  = vqtbl1q_u8(c2s_vec, codes);
// Save 16 symbols at once
vst1q_u8(symbols + i, syms);
```

See how we can decode 16 symbols with just bit-unpacking and 3 extra instructions.

### 3.3. Bottom-up primitive performance

Table 9 demonstrates bottom-up primitive-performance. Looking at the **ns/elem** metric, we see that all primitives achieve performance comparable or better to the fast `partition` primitives from Table 4, and none pay the memory-overload penalty that the slow `scatter` primitives suffered from.

### 3.4. Bottom-up decoding performance

To evaluate the performance of bottom-up PivCo-Huffman decoding we looked at our datasets on two machines. We're also testing the impact of tree-complexity optimizations from Section 2.2. Table 10 and Figure 11 show the results. We can see that PivCo-Huffman decisively beats decoding performance of Huff0 and Oodle Huffman on all datasets and platforms. The magnitude of the PivCo-Huffman benefits depends on three factors:

| Dataset | M4 | | | | | c8i | | | | |
|---|---|---|---|---|---|---|---|---|---|---|
| | PivCo-Huffman tree opt. | | | Huff0 | Oo-Huff | PivCo-Huffman tree opt. | | | Huff0 | Oo-Huff |
| | naive | flat | flat-opt | | | naive | flat | flat-opt | | |
| **proba80** | 16021 | 17564 | **17476** | 2929 | 3343 | 37789 | 38224 | **37854** | 1936 | 2131 |
| **english** | 5033 | 6059 | **6981** | 2747 | 3293 | 6272 | 7904 | **10738** | 1882 | 2123 |
| **html_wiki** | 2815 | 4337 | **4730** | 2269 | 3262 | 1957 | 4075 | **4808** | 1563 | 2107 |
| **prose_pride** | 3719 | 4820 | **5174** | 2547 | 3158 | 2687 | 4987 | **5720** | 1758 | 2120 |
| **json_api** | 3244 | 4137 | **4484** | 2336 | 3233 | 2270 | 4131 | **5022** | 1613 | 2121 |
| **dna_fasta** | 10345 | 10537 | **10150** | 2789 | 3227 | 20571 | 21640 | **22186** | 1927 | 2132 |
| **chinese_text** | 2789 | 3989 | **4913** | 2235 | 3232 | 2116 | 4006 | **5384** | 1550 | 2112 |
| **calgary_pic** | 6667 | 10513 | **10751** | 2491 | 3257 | 3420 | 8223 | **9890** | 1676 | 2105 |

Table 10: PivCo-Huffman (bottom-up) decode bandwidth (MB/s) for different tree optimization levels. We compare naive, flat-subtrees and optimized flat-subtrees against Huff0 and Oodle Huffman.

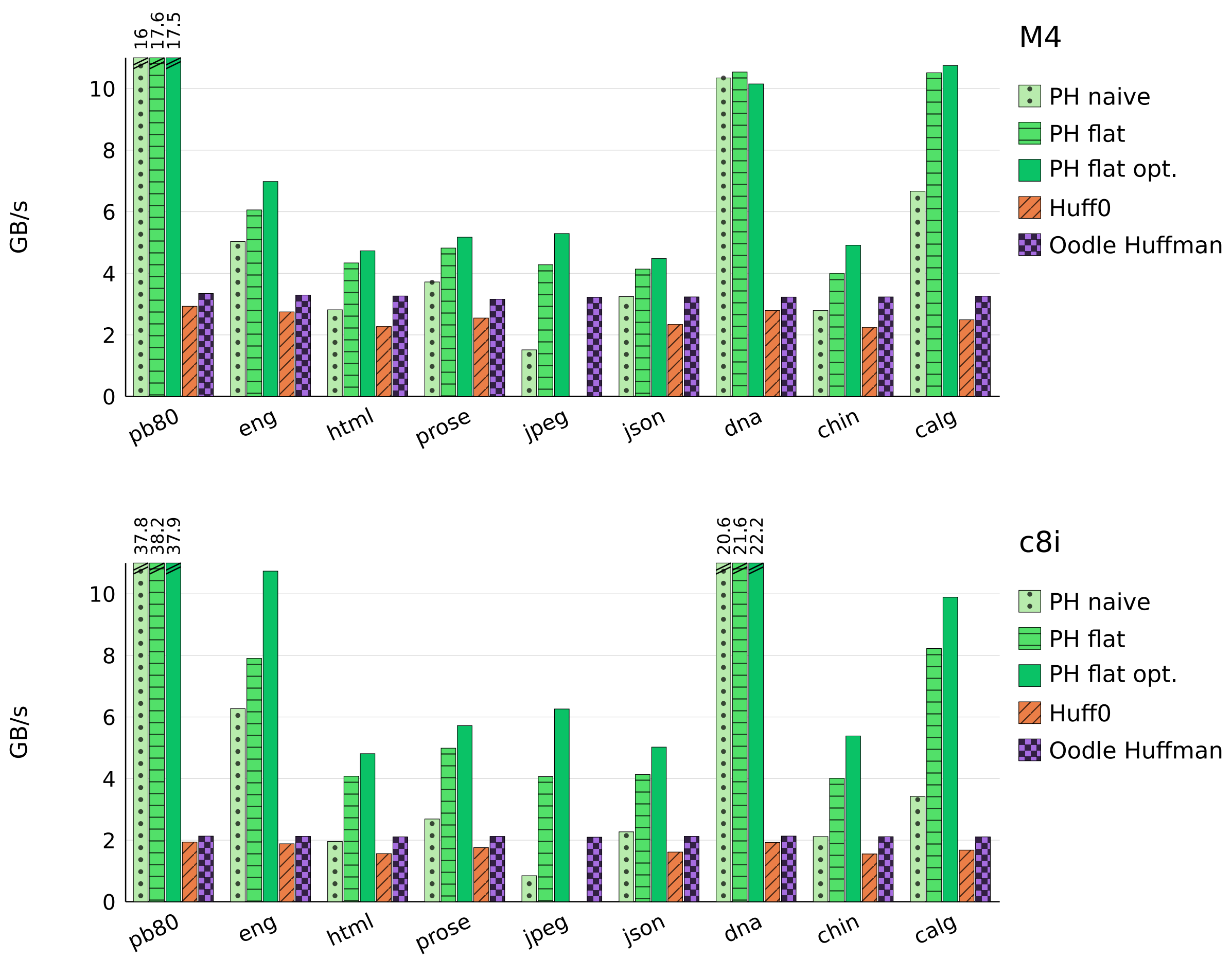

Figure 11: PivCo-Huffman decoding performance with different tree optimization levels

- tree level optimization - we see that both flat subtrees and their optimized versions provide significant benefits. This makes sense, as with the reduction of the number of operations, the performance improves.

- dataset - skewed datasets benefit most, as on these PivCo-Huffman can reduce the number of operations for shorter codes / more frequent symbols. In particular, tree optimizations have no impact on *proba80* and only marginal on *dna_fasta*.
- CPU - M4 provides great OoO scalar evaluation, helping traditional algorithms more. On the other hand, c8i has weaker scalar processing, but more performant SIMD - with AVX-512 - this benefits PivCo-Huffman. As a result, while PivCo-Huffman is consistently better on M4, that difference is even more pronounced on c8i.

# 4. Encoding

Data encoding for PivCo-Huffman is relatively straightforward - it, naturally, uses the same tree shape that we build for decoding, and follows the same pattern of tree traversal with high-performance SIMD primitives.

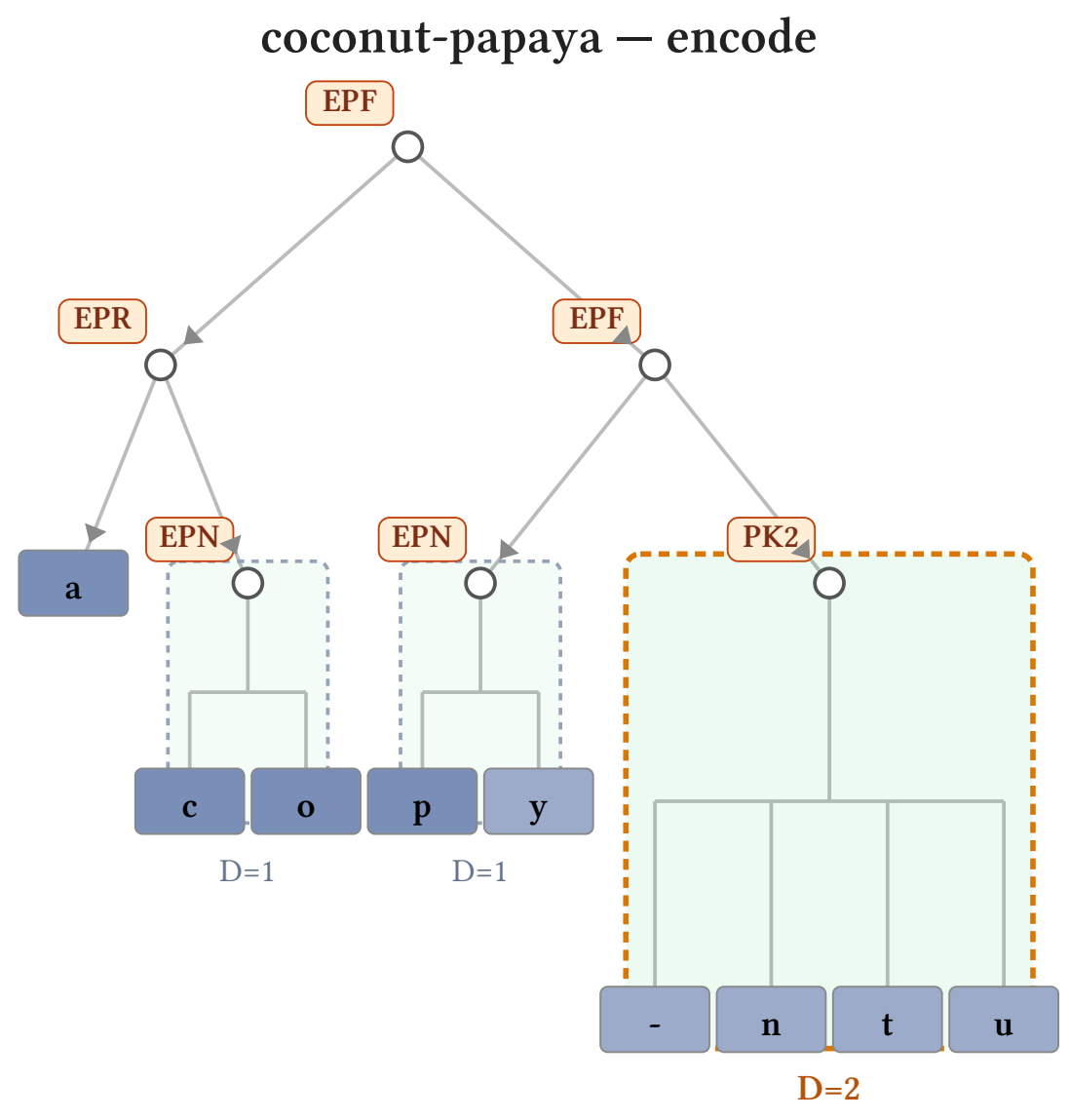

Figure 12: Encoding tree operations (*optimized flat trees* off)

| Symbol | Explanation |
|---|---|
| EPF | `enc_partition_full` - create a bitmap, split codes into left/right outputs |
| EPL | `enc_partition_left` - right child is a leaf, only produce bitmap+left codes |
| EPR | `enc_partition_right` - left child is a leaf, only produce bitmap+right codes |
| EPN | `enc_partition_none` - both children are leaves, only produce bitmap |
| PKN | `packN` - pack codes into N-bits sequence, used for flat subtrees |

Table 11: Primitives used in encoding

Figure 12 shows operations for an example encoding tree, and Table 11 lists operations used in that phase. Note extreme similarity to Table 1 from Section 2. In fact, `enc_partition_*` operations are functionally equivalent to first building a bitmap from symbols, and then applying the `partition` operations used in top-down decoding. A small difference is that the elements we partition are not 16-bit *indices in the output*, but 16-bit *codes*. Still, the primitives are the same.

Table 12 and Figure 13 show the encoding performance on various datasets and hosts. End-to-end results show a fair comparison with other solutions. PivCo-Huffman's performance is hindered here by the Huffman-code creation time, dominated by symbol frequency counting. The "prebuilt-tree" results show the actual encoding performance, showing e.g. how highly-skewed datasets can achieve very high "raw" encoding performance.

The final compressed data consists of the Huffman codes information, followed by the per-node information in a deterministic tree-traversal order. This information, including byte-padding overheads, is included in the compression-ratio results. Appendix D provides the complete wire-format layout.

| Dataset | PivCo-Huffman | | | | Huff0 | | Oo-Huff | |
|---|---|---|---|---|---|---|---|---|
| | end-to-end | | prebuilt tree | | | | | |
| | M4 | c8i | M4 | c8i | M4 | c8i | M4 | c8i |
| proba80 | 1477 | 1678 | 3565 | 8675 | 1334 | 1147 | 1740 | 1295 |
| english | 1459 | 1615 | 2189 | 3290 | 1771 | 1309 | 2520 | 1443 |
| html_wiki | 1231 | 1110 | 1750 | 2186 | 1766 | 1260 | 2344 | 1248 |
| prose_pride | 1240 | 1258 | 1809 | 2584 | 1722 | 1267 | 2325 | 1364 |
| image_jpeg | 1292 | 1114 | 2176 | 3026 | na | na | 2688 | 1380 |
| json_api | 1200 | 1172 | 1737 | 2324 | 1753 | 1303 | 2359 | 1365 |
| dna_fasta | 1413 | 1826 | 2513 | 5647 | 1550 | 1308 | 2144 | 1449 |
| chinese_text | 1250 | 1195 | 1760 | 2287 | 1823 | 1277 | 2461 | 1332 |
| calgary_pic | 1261 | 1190 | 3111 | 5671 | 1297 | 1063 | 2283 | 1180 |

Table 12: Encoding performance on M4 and c8i (MB/s). For PivCo-Huffman, we report both "end-to-end" and "prebuilt tree" results

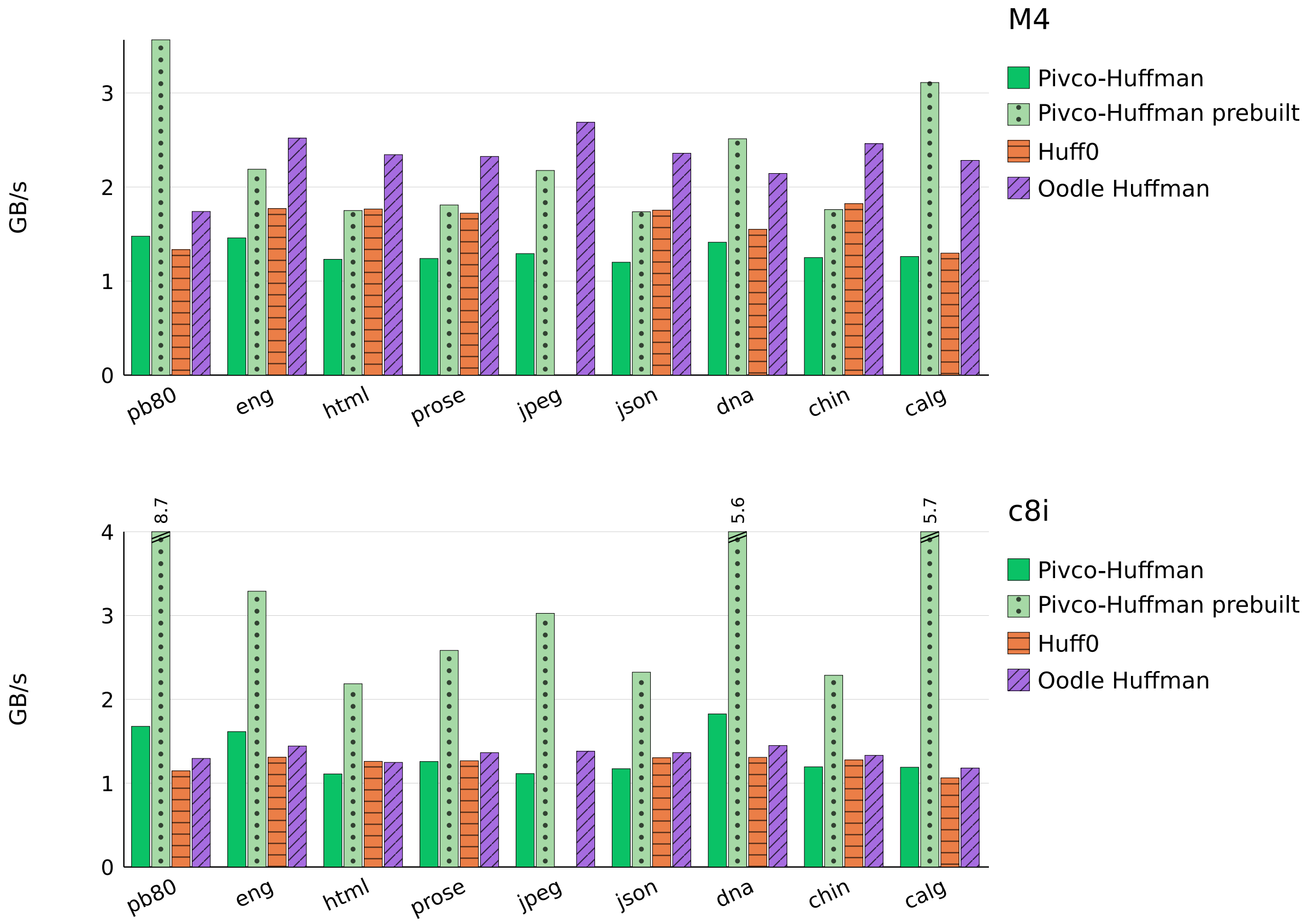

Figure 13: Encoding throughput, M4 (top) and c8i (bottom) — the data of Table 12. PivCo-Huffman end-to-end (`ph-op`) and prebuilt-tree (`ph-pb`) vs Huff0 and Oodle Huffman. On c8i the `ph-pb` proba80 bar exceeds the axis cap and is clipped (true value labeled).

## 5. Breaking the bit-barrier

Huffman encoding, while ubiquitous, has one key limitation: its code lengths are constrained to whole bits. That means, for some distributions, it is further from entropy-optimal than desired.

| Distribution | $H$ (bits) | Huffman (bits) | $H$ / $Huff$ | $Huff$ - $H$ | max node benefit |
|---|---|---|---|---|---|
| **calgary_pic** | **1.210** | **1.690** | **0.716** | **0.480** | **0.446** |
| chinese_text | 5.814 | 5.978 | 0.973 | 0.163 | 0.007 |
| **dna_fasta** | **2.080** | **2.265** | **0.918** | **0.185** | **0.171** |
| english | 4.225 | 4.247 | 0.995 | 0.022 | 0.006 |
| html_wiki | 5.476 | 5.634 | 0.972 | 0.158 | 0.009 |
| image_jpeg | 7.887 | 7.917 | 0.996 | 0.031 | 0.008 |
| json_api | 5.198 | 5.240 | 0.992 | 0.041 | 0.006 |
| **proba80** | **0.904** | **1.250** | **0.723** | **0.346** | **0.278** |
| prose_pride | 4.529 | 4.605 | 0.984 | 0.076 | 0.013 |

Table 13: Per-dataset entropy gap and peak single-node bitmap benefit (bits/byte). Bold rows are the heavily-skewed datasets. In all of these, a single partition node captures most of the Huffman redundancy.

A well-known solution to this problem is arithmetic coding ([3]); however, it has not been popular due to its performance and patent controversies. Luckily, Jarek Duda proposed *ANS-based encoding* ([4]), which solves both problems. It led to two main algorithms: tabled-ANS (**tANS**) and range-ANS (**rANS** [14]), both allowing codes to have lengths close to entropy-optimal.

We will focus on tANS, which is used by FSE ([8]) and Oodle TANS library ([15]), both related to the Huffman implementations we discussed in Section 1.2.

The typical tANS decoding is actually quite similar to an optimized Huffman decoding routine from Section 1.2:

```
t = decoding_table[state];
state = t.newX + read_bits(t.numBits);  //state transition
emit_symbol(t.symbol);                  //decoded symbol
```

While similar, the critical difference is that we see a `state` variable which is carried through the iterations and mutated depending on the data. Also, the `decoding_table` is constructed such that, for the same symbol, a different number of bits may be consumed depending on the current `state`.

As a result, the same approach to tANS decoding as we did to Huffman in Section 2 is not directly applicable. Still, we will demonstrate how PivCo-Huffman opens a unique opportunity to apply ANS.

### 5.1. Skew analysis in Huffman trees

Table 13 shows additional analysis for datasets from Appendix A. We can see that for most of them, Huffman encoding actually achieves almost perfect code length, reaching typically 97-99% of entropy. As a result, for most of them, applying a more expensive compression method is probably not useful. Three datasets stand out:

- **proba80** - artificial dataset, skewed on purpose
- **calgary_pic** - a mostly-white bitmap
- **dna_fasta** - DNA dataset, mostly **`"A C G T"`** letters plus some extras

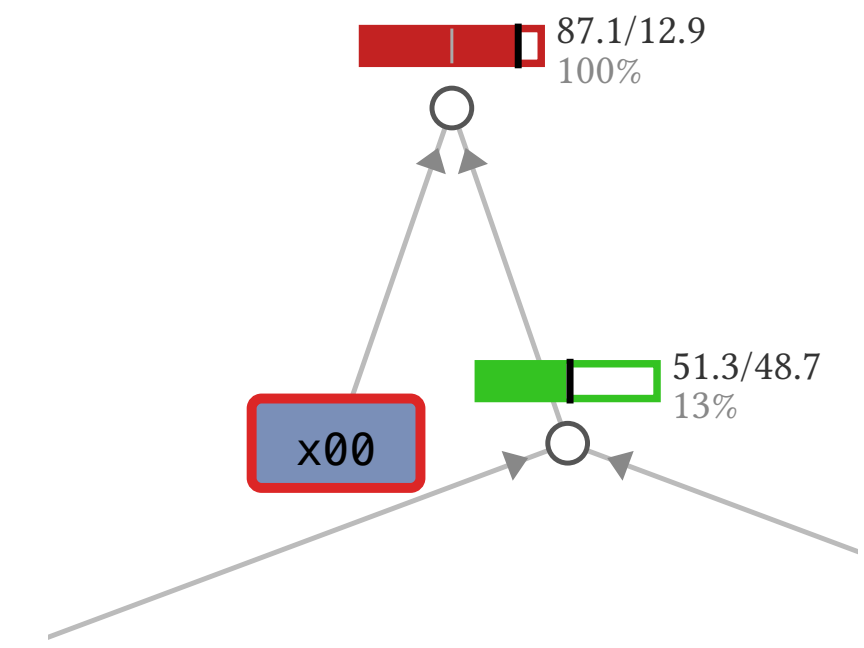

Figure 14: Skew visualization for **calgary_pic** (top part of the tree only)

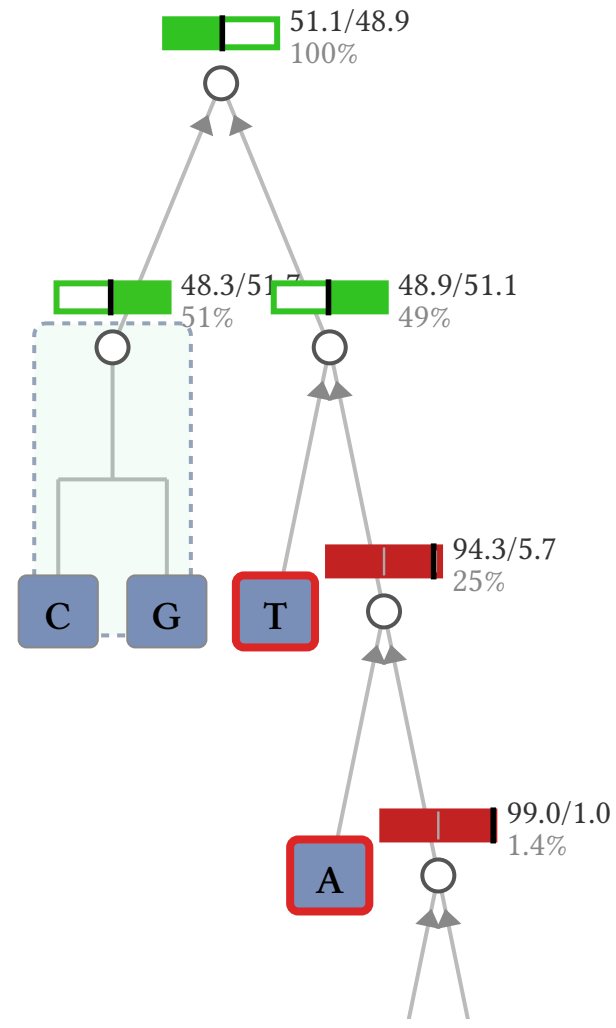

Figure 15: Skew visualization for **dna_fasta** (top part of the tree only)

Figure 14 and Figure 15 show visualization of the top parts of the tree for **calgary_pic** and **dna_fasta** datasets. For each tree node, we report left/right skew, and the percentage of data covered by a given subtree. Bar color reflects skew severity (red - highly skewed).

For **calgary_pic** we see how the root node has a skew of 87.1/12.9, with **H=0.554** That means, if that **one node** was entropy-encoded, we would save **0.446** bits per encoded element, almost reaching the Huffman encoding gap of **0.480**.

**dna_fasta** is slightly different, because the interesting node is not the root, but a node 2-levels deep. With **`"A C G T"`** symbols occupying the vast majority of the input, **`"C G T"`** were assigned 2-bit codes, but **`"A"`** had to be assigned a 3-bit code to make room for the remaining, infrequent symbols. As a result, 25% of all symbols get to the parent node of **`"A"`** and 94.3% of those go to **`"A"`**. That node has **H=0.315**, so entropy-encoding would save **0.685** bits for each symbol, but with only 25% of the input reaching that node, it results in **0.171** average bits saved per code, also very close to the **0.185** bits Huffman gap. Note that the sibling node of **`"A"`** has an even stronger skew, but with only 1.4% of data reaching it, optimizing it is not worth it.

### 5.2. PivCo-Huffman+ANS implementation

The analysis above suggests that for most datasets applying ANS-based encoding is not worth the additional complexity. This is consistent with what e.g. [8] does - the literal stream is only Huffman compressed, but the significantly skewed length/offset data is tANS-compressed.

Additionally, we see how for datasets where ANS-encoding *would* be useful, the vast majority of benefit often comes from just a few (usually one, sometimes two) nodes in the Huffman tree. To exploit that, PivCo-Huffman was extended with *selective ANS encoding — applied only to the nodes where it matters*. This means that for most datasets no ANS overhead is paid, and when it is applied, it is only paid for a small subset of the data.

A concrete implementation of which node should be FSE-selected is currently as follows:

- node needs to have at least `PIVCO_FSE_MIN_BITMAP_BYTES` (32 bytes/ 256 bits default)
- node skew needs to be higher than `PIVCO_FSE_MIN_THRESHOLD` (0.625 default)
- fse *benefit* needs to be better than `PIVCO_FSE_MIN_RATIO` (0.95 default) with benefit computed as `(depth + fse_H) / (depth + 1)`, where **fse_H** is the average bit-cost of fse-encoding (close to *H*, but usually a bit higher). The motivation here is that while saving e.g. 0.2 bits for a root-node makes sense, doing it for a node at depth 5 (so already 5-bits long) is probably not worth it.

Note that we know all of the above information purely from the symbol frequencies used to construct the Huffman tree; we do not need to gather any additional data statistics. We *know* when ANS will help.

When we decide to compress a particular bitmap with ANS, today we use FSE ([8]). Note that we compress the bitmap as *bytes*, not as bits. This means that for each symbol decoded with FSE, we cover *8 symbols* of the original bitmap. This, combined with applying FSE selectively, is critical to making PivCo-Huffman+ANS efficient.

One non-trivial cost of FSE is creation of decode tables. To avoid it, we use statically precomputed 50 decode tables for bitmap skew in range (50,51,..,98,99)%. Then, we simply choose a table based on the symbol skew during encoding/decoding.

An interesting aspect of this decision is that the tables above are built for *bytes* constructed using a random *bit* distribution. Depending on the actual distribution of bits, this can result in compression efficiency lower than if we actually built the table for a specific dataset. For example, let us take a collection of 300 `0xFF` and 100 `0x00` values. If the decoding table were custom-built, these values would take 75% and 25%, respectively, of the frequencies. However, since the pre-built partitions assume random bit distribution, these symbols' expected frequencies will be much lower, resulting in more bits assigned to them during encoding. This shows that our approach might not be able to exploit some order-based compression ratio opportunities in the data.

Note, we use a *tuned* version of FSE (*x8y1*), as we found that the default implementation can be significantly improved for our needs, see Appendix F. See also Appendix E.4 for another possible optimization.

### 5.3. Benefits

This approach to FSE application in PivCo-Huffman+ANS has the following benefits:

- FSE is slower than Huffman, but since each FSE-symbol we decode covers 8 Huffman-symbols from our main tree, we pay only 1/8th of the cost per bitmap.
- It can be applied *only* to nodes where it actually matters (mostly highly skewed)
- Compression ratio vs performance can actually be *tuned* (slightly) depending on the actual FSE-triggering strategy
- There is no FSE table construction
- The FSE table selection can be done for every decompression block separately (8KB). This allows exploiting locally-optimum distributions. Stock FSE decides on the decoding table every 128KB, and so it will not exploit these local properties.

### 5.4. Results

Table 14 compares the PivCo-Huffman (**PH**) and PivCo-Huffman+ANS (**PHA**) performance with Huff0 and Oodle's TANS library. Figure 16 visualizes these results, additionally including FSE. We see that for non-skewed datasets, **PH** and **PHA** achieve the same performance, but for skewed datasets **PHA** detects an opportunity to *selectively* apply FSE - this brings the compression ratio close to full FSE, while still achieving significantly higher decode performance. Interestingly, for *dna_fasta* we see that the FSE performance impact is relatively low, as FSE is applied to a non-root node covering only 25% of the symbols. Finally, the compression ratio of the *calgary* dataset (a scanned text-on-white image) showcases the impact of PivCo-Huffman+ANS utilizing locally-optimal decompression tables.[3]

[3]The author by no means suggests PivCo-Huffman+ANS is better than FSE - it just occasionally has this slightly unexpected property

| Dataset | PH | | | PH+ANS | | | Huff0 | | | oo-tans | | |
|---|---|---|---|---|---|---|---|---|---|---|---|---|
| | ratio | M4 MB/s | c8i MB/s | **ratio** | **M4 MB/s** | **c8i MB/s** | ratio | M4 MB/s | c8i MB/s | ratio | M4 MB/s | c8i MB/s |
| *proba80* | *6.29* | *17476* | *37854* | ***8.45*** | ***6566*** | ***5353*** | *6.40* | *2929* | *1936* | *8.84* | *2596* | *1469* |
| english | 1.86 | 6981 | 10738 | **1.86** | **6963** | **10777** | 1.88 | 2747 | 1882 | 1.89 | 2553 | 1448 |
| html_wiki | 1.45 | 4730 | 4808 | **1.46** | **4682** | **4594** | 1.47 | 2269 | 1563 | 1.48 | 2501 | 1418 |
| prose_pride | 1.72 | 5174 | 5720 | **1.72** | **5068** | **5622** | 1.75 | 2547 | 1758 | 1.76 | 2521 | 1420 |
| image_jpeg | 1.00 | 5291 | 6260 | **1.01** | **4728** | **5469** | — | — | — | 1.01 | 2463 | 1385 |
| json_api | 1.51 | 4484 | 5022 | **1.51** | **4417** | **5053** | 1.53 | 2336 | 1613 | 1.54 | 2493 | 1413 |
| *dna_fasta* | *3.52* | *10150* | *22186* | ***3.79*** | ***6496*** | ***7995*** | *3.55* | *2789* | *1927* | *3.83* | *2530* | *1449* |
| chinese_text | 1.35 | 4913 | 5384 | **1.35** | **4913** | **5108** | 1.37 | 2235 | 1550 | 1.38 | 2462 | 1406 |
| *calgary_pic* | *4.64* | *10751* | *9890* | ***6.95*** | ***4139*** | ***3467*** | *4.80* | *2491* | *1676* | *6.55* | *2520* | *1443* |

Table 14: PivCo-Huffman+ANS benchmark: compression ratio (higher = better) and decode throughput (MB/s) on M4 and c8i. **PH** and Huff0 are plain Huffman (≈equal ratio); **PHA** improves ratio from ANS-coded partition bitmaps. Skew-heavy datasets in *italic*. "Calgary" compression ratio in red. Huff0 refuses to compress the *image_jpeg* dataset.

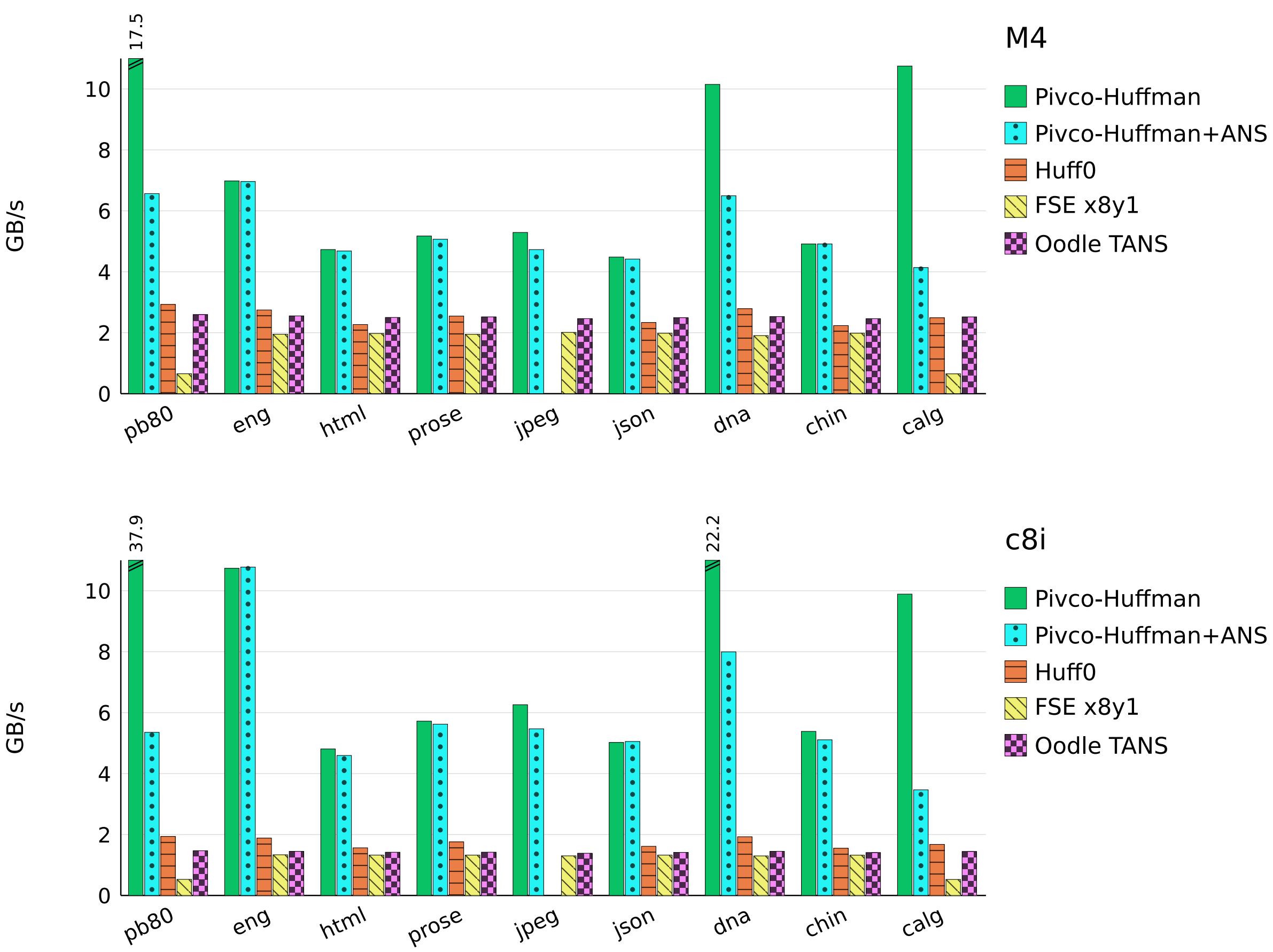

Figure 16: Decode throughput per engine, M4 (top) and c8i (bottom) — the bandwidth columns of Table 14, with the addition of **FSE x8y1**

# 6. Challenges and opportunities

While PivCo-Huffman is an interesting approach, the author does not claim it's a drop-in replacement for existing solutions. In this section we discuss the challenges it faces, but also further research and improvement opportunities.

## 6.1. Data size

PivCo-Huffman performs a tree traversal, with the tree size comparable to the dictionary size (minus the flat-subtree optimization). Additionally, on each level, it requires a sizable amount of data to achieve high performance (typically 64+ codes). That means, that a minimum data size at which its application makes sense is in the range of at least kilobytes.

## 6.2. SIMD-focus

To achieve high performance, PivCo-Huffman requires a careful design of its primitives using SIMD. As a result, it might be challenging or even impossible for it to achieve satisfying performance on hardware platforms or in programming languages that do not provide such capabilities.

On the flip side, branch-free, SIMD-friendly code in PivCo-Huffman suggests this approach might perform well on GPUs.

## 6.3. Primitive optimization

While presented primitives already provide good performance, optimizations are surely possible, especially when targeting different hardware.

For example, we prototyped significantly faster `merge-flat-D2/D3` primitives for AVX-512 that use bit-sliced packing to achieve around 1.5x performance. However, this version required a wire format change, which resulted in slower performance on other systems. Still, optimizations like that are feasible in specific scenarios.

While this is an opportunity for further improvement, it is also a challenge, as managing multiple primitives for different backends can be complex. Additionally, if we strive for optimal performance, the best implementation can be specific not only to the ISA, but even to the actual CPU used. For example, we have seen optimizations that provided benefits on M4, but reduced performance on Graviton.

## 6.4. Exploring bitmap-compression alternatives

While per-bitmap FSE is a promising optimization, other strategies are possible. Alternative bitmap-compression methods could provide an interesting design point in a performance/compression ratio space.

## 6.5. Impact on LZ-codecs

Huffman and other entropy-codecs are rarely applied standalone; they are most commonly used as a step after LZ-family or similar compressors. We analyzed PivCo-Huffman's behavior on the data streams generated by zstd ([9]) after the LZ-style compression pass:

- literal stream (Huffman in zstd) - PivCo-Huffman achieves comparable compression ratio at higher speed. The skew present before LZ is typically significantly reduced, making ANS applicability here much lower.
- literal-length stream (FSE in zstd) - highly skewed; PivCo-Huffman+ANS provides compression ratios close to FSE at a much higher speed
- match-length and offset (FSE in zstd) - less skewed than LL; limited applicability of ANS.

In our experiments FSE often took more than 50% of the decode time, suggesting that an implementation using PivCo-Huffman+ANS instead *could* be an interesting point in the size/speed space. At the same time, due to PivCo-Huffman's much higher implementation complexity than that of e.g. zstd, the author does not believe it is a good building block for a general-purpose codec.

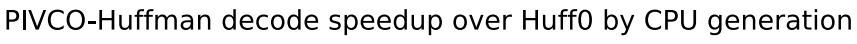

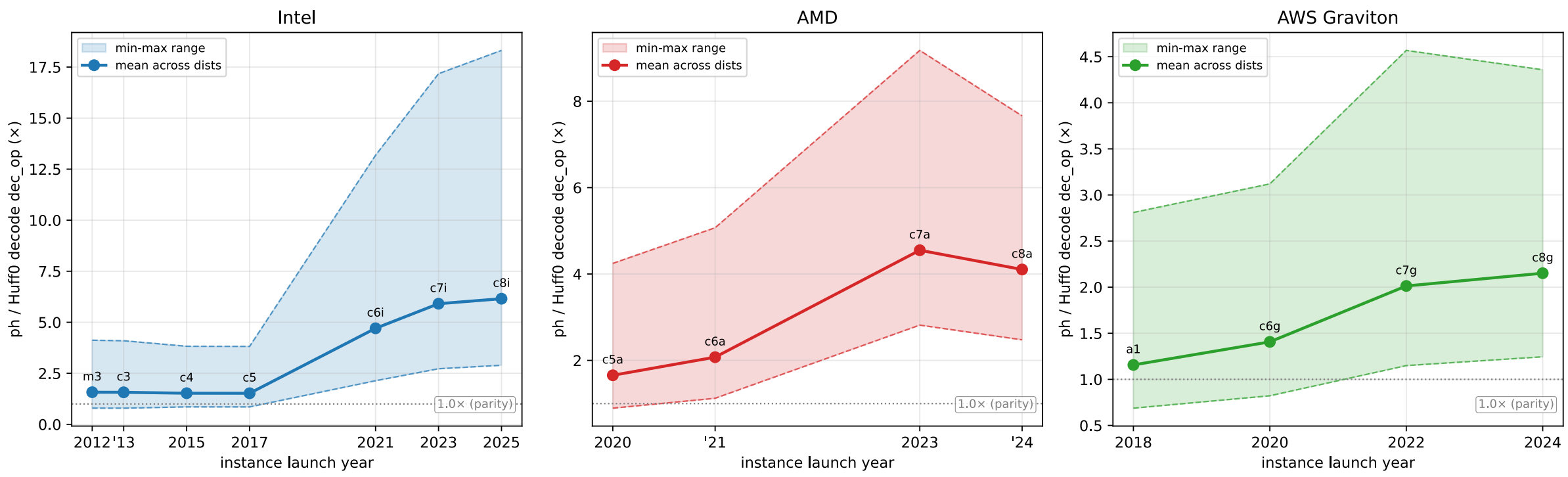

Figure 17: PivCo-Huffman performance over Huff0 on 3 CPU families on AWS (across datasets from Appendix A)

### 6.6. CPU development trends

While the presented PivCo-Huffman performance focuses on recent CPUs, it is interesting to see how it performs on older hardware. Figure 17 shows that PivCo-Huffman provides consistent benefits on all these architectures, even going back to 2012, but the benefit *increases* with newer CPU generations. The main reasons are the improvement in SIMD capabilities and performance, and the fact that PivCo-Huffman presents a lot of simple, predictable work that utilizes modern wide out-of-order CPUs well.

The trend is strongest on Intel — the Ice Lake generation (2021) introduced the AVX-512 VBMI2 family, which provides the byte-granular `vpcompressb` instruction that maps very directly to PivCo-Huffman's partition kernel. AMD followed with Zen 4 in 2023. ARM does not currently have an equivalent instruction — even at 256-bit SVE width ([16]), `svcompact` operates only on 32- and 64-bit elements, which we measured to be slower than NEON's compress-table approach. A future ARM ISA with such an instruction would likely produce a similar step-function for PivCo-Huffman on Graviton-class hardware.

## 7. Related work

### 7.1. Entropy coding

Entropy-based encoding systems have been intensely researched for many decades, with Huffman, arithmetic compression and (recently) ANS-family being the most popular, both in research and in applications. Other well-known approaches include Golomb-Rice ([17], [18]), Elias-Fano ([19]), Tunstall ([20]), lightweight integer compression ([21], [22]), and others ([23]).

In this context, PivCo-Huffman can be seen as a performance-focused variant of Huffman. The ANS application, while useful, is more of an extension to this core idea.

Separately, it would be interesting to see if some of the techniques applied in this paper could be used to other methods in this space. For example, Appendix G discusses how the *pivoted coding* approach can be applied to Golomb coding.

### 7.2. Wavelet trees

Wavelet trees, introduced in [12], are a popular structure used in many different applications, typically succinct indexing, full-text indexing, and even compression (see [24]).

| Dimension | Wavelet Trees | PivCo-Huffman |
|---|---|---|
| **Core representation** | Alphabet tree with node bitmaps | Code tree with node bitmaps |
| **Primary purpose** | Indexed sequence representation: access, rank, select, range queries, etc. | Sequential compression/decompression throughput |
| **Aux structures** | Usually add rank/select support per bitmap | none |
| **Operations** | Navigate query positions through levels | Reconstruct whole dense output stream |
| **Node bitmap constraints** | Often must remain rank/select-friendly, e.g. use RRR [26] | Can use decode-friendly encodings, including FSE/ANS |
| **Tree shape** | Fixed/balanced, Huffman-shaped, wavelet matrix variants, etc. | Huffman-derived with flat subtrees |
| **Performance target** | Query latency/space tradeoff | GB/s-scale sequential decode throughput |
| **Block model** | Often whole sequence/static text index | Block codec, streaming possible |

Table 15: Comparison of wavelet trees and PivCo-Huffman

PivCo-Huffman reuses the idea of a "tree of bitmaps" from wavelet trees, but to the author's knowledge, most other aspects of the solutions are quite different; see Table 15 for comparison. Still, there is definitely some interesting overlap, especially around wavelet-tree creation, suggesting that ideas from wavelet-trees research could be applied to PivCo-Huffman and the other way around. For example, [25] proposes a *bottom-up building* of wavelet trees, and [13] apply SIMD instructions to this problem.

### 7.3. Bit-packing

*Flat-subtrees* are one of the key performance aspects of PivCo-Huffman. Their implementation depends heavily on packing/unpacking D-bit integers, often called a *Frame-of-Reference* coding (which additionally applies an offset). This problem appears in many different areas, including databases and information retrieval. A lot of work focuses on this problem in its original setting, where values are packed contiguously (e.g. [21], [27]). However, other approaches use non-linear data organization allowing them to achieve much higher performance ([28], [29]).

PivCo-Huffman currently uses a relatively straightforward, linear, SIMD-based bit-packing. Applying techniques from other work in that space could possibly further improve PivCo-Huffman's performance.

## 8. Conclusions

In this paper we presented PivCo-Huffman, a novel approach to Huffman-compression, based on the structure from *wavelet trees*. The main contributions are:

- adapting the bitmap format of *Huffman-shaped wavelet trees* into a sequential block-compression wire format
- compression-focused tree-structure optimizations allowing better encoding/decoding performance
- novel "bottom-up" decoding approach, eliminating the *scatter* problem of the more natural top-down approach
- highly performant SIMD primitives for both encoding and decoding operations
- the concept of *selective* application of ANS-based encoding in the Huffman tree to further improve its compression ratio

The resulting approach, while slightly slower on the encoding side, in our tests consistently beats the decoding speeds of the state-of-the-art solutions by a large margin. It also provides compression ratios approaching the ANS-based methods at significantly better speeds.

### 8.1. AI disclosure

Anthropic Claude was used extensively during development of this project, especially in areas like coding, testing, automation and research. It also contributed many small ideas and improvements, especially around SIMD code. At the same time, the author declares that the vast majority of the key ideas and concepts here are human-invented. This document is purely human-written (except for spellcheck etc.).

### 8.2. Acknowledgments

The author would like to thank Fabian "Ryg" Giesen for his help with Oodle and his comments on an early draft of the paper.

# Appendices

## A Datasets

| Name | #syms | $H$ (bits) | Huffman (bits) | min | max | Source / description |
|---|---|---|---|---|---|---|
| **calgary_pic** | 159 | 1.210 | 1.690 | 1 | 11 | Calgary Corpus 1bpp CCITT scanned page, proba80-like real data (calgary_pic) |
| **chinese_text** | 151 | 5.814 | 5.978 | 4 | 11 | Project Gutenberg Hong Lou Meng, Chinese UTF-8, truncated (chinese_text.txt) |
| **dna_fasta** | 38 | 2.080 | 2.265 | 2 | 10 | NCBI E. coli K-12 genome FASTA, truncated ~500 KB (dna_fasta.fa) |
| **english** | 30 | 4.225 | 4.247 | 3 | 10 | Synthetic: English letter relative frequencies |
| **html_wiki** | 197 | 5.476 | 5.634 | 4 | 11 | English Wikipedia "Cat" article, served HTML (cat-wiki.html) |
| **image_jpeg** | 256 | 7.887 | 7.917 | 6 | 10 | Wikimedia Commons JPEG photo Cat03.jpg (near-uniform 256 bytes) |
| **json_api** | 98 | 5.198 | 5.240 | 3 | 11 | GitHub API commits feed JSON (json_api.json) |
| **proba80** | 6 | 0.904 | 1.250 | 1 | 5 | Synthetic FSE benchmark: geometric, top symbol p~0.80 (very skewed) |
| **prose_pride** | 96 | 4.529 | 4.605 | 3 | 11 | Project Gutenberg plain-text Pride and Prejudice (pride.txt) |

Table 16: Used datasets: alphabet size, entropy $H$, mean Huffman code length, code-length min/max, and source. Available in PivCo-Huffman repo

## B Machines Tested

| Symbol | Machine type | Arch | CPU family | CPU year |
|---|---|---|---|---|
| **M4** | Apple MacBook Pro (Mac16,6) | aarch64 | Apple M4 Max | 2024 |
| **c8i** | AWS EC2 c8i.large | x86-64 | Intel Xeon 6 (Granite Rapids) | 2024 |

## C Testing methodology

In our testing we use datasets from Appendix A and machines from Appendix B.

Unless stated otherwise, we compute the time that *includes* the setup time.

For datasets that are smaller than 1MB, we create copies to cross the 1MB size.

For bandwidth tests, to determine the optimal performance of each algorithm (reduce noise etc), we use the following setup:

- one *run* performs 20 repetitions of the operation back-to-back
- we execute 20 *runs* and collect the timings
- if the top-two *run* timings are within 2%, we stop
- otherwise, two more *runs*, until the 2% difference goal is met, up to 40 times in total.

- the best *run* time is used

# D Compressed data organization

There are three main components to storing data compressed by PivCo-Huffman: symbol codes, compressed data, and metadata.

For symbol codes, we simply store 128 bytes containing the canonical Huffman-code lengths (256 times 4 bits). Note, that today we limit the Huffman code lengths to 11 bits (similarly to Huff0). For many datasets this might be suboptimal, but since this part is a tiny percentage of the compressed data, we keep it simple. From this data, we can reconstruct the canonical Huffman codes, from which we can construct the actual exact PivCo-Huffman tree.

The actual compressed data is stored in *per-8KB blocks* (configurable). It starts with a 4-byte block compressed-size information, followed by the per-node data, in the tree-traversal order.

For the *internal nodes*, we store:
- 2-byte (optional) symbol count of the right child - necessary to find its data in the stream. Only used for nodes with both non-leaf children.
- 1-byte FSE marker:
  - `0` means no FSE is applied
  - lower 7 bits - the FSE table id (starting at 1)
  - top bit - *XOR marker* - if set, it means that the FSE-compressed data is on *reversed* input bits. This is used when the skew in the bitmap is in the opposite direction to what the table is built for.
- bitmap body
  - if FSE marker is `0`, we store `ceil(n/8)` bytes.
  - otherwise, we store a 2-byte length of FSE-compressed data followed by the FSE stream.

For *flat subtree roots*, we simply store `ceil(n*D/8)` bytes containing the compressed data.

The final component is file-level metadata which includes total uncompressed size, checksums, block size and other necessary fields. We use a simple file format containing this information, allowing easy testing of PivCo-Huffman.

# E Failed optimizations

As we worked on this paper, we have tried many things that didn't pan out. We list them briefly in here for the reader, either to save some time, or perhaps inspire to try harder.

## E.1 Root-levels decoding

For top-down decoding, before we settled on flat-subtrees, we investigated the idea of decoding the top D-deep part of the tree for situations where the shortest code was D-bits long. The intuition was that in one operation we would cover a lot of the most frequent nodes. However, such an operation would result in 2^D-way stream partitioning, which turned out to be simply too slow.

## E.2 Fusing `scatter` and `partition`

When we tried to optimize top-down decoding, we realized that `scatter` and `partition` are limited by different CPU resources - `scatter` by writes, and `partition` by table lookups and computations. We tried to combine them, by having `scatter` for one decompressed block also perform a part of the `partition` effort for the following block. Alas, we couldn't achieve any significant benefits.

## E.3 Tree optimizations for FSE

Similarly to optimizing the Huffman tree to reduce the number of operations (see Section 2.2), we tried optimizing the tree to maximize the benefit of FSE.

Two approaches have been attempted:

- allowing *splitting* of the *flat trees* if the root node had significant skew
- arranging a tree in a left-heavy (by frequency) way, to force more "skewed" nodes

While both optimizations provided occasional benefits, the impact was so small we decided to park them, especially as both required transferring the actual frequencies (not only code lengths) to the decompressor.

### E.4 Fusing FSE with `merge`

To further optimize PHA performance, we tried to fuse the FSE decoding with the `merge` step of the bottom-up processing. While we achieved small improvements (a few percent), the complexity of this solution was not worth incorporating into the code base.

## F Tuning FSE

When working with FSE for Section 5, we noticed that the FSE overhead had a more severe impact than we would like. As a result, we performed a side experiment where we tuned FSE's main loop. By default, it looks like this (slightly simplified):

```
while ((BIT_reloadDStream(&bitD) == BIT_DStream_unfinished)   // reload bits
        & (op + 4 <= olim)) {
    op[0] = FSE_decodeSymbolFast(&states[0], &bitD);
    op[1] = FSE_decodeSymbolFast(&states[1], &bitD);
    op[2] = FSE_decodeSymbolFast(&states[0], &bitD);
    op[3] = FSE_decodeSymbolFast(&states[1], &bitD);
    op += 4;
}
```

In that code, `states` refers to a table of two states in the FSE table - this is similar to using two independent cursors and provides more independent instructions to modern CPUs. Still, the data for both states comes from a single, interleaved stream. We also see that the loop is explicitly *2-unrolled*, reducing the loop overhead. We call this particular implementation **x2y2** (x: 2 cursors, y: 2-unroll).

We performed a thorough testing of equivalent implementations of FSE with **x={2,4,6,8,10,12,16}** and **y={1,2,4}** on a number of machines. The example results for M4 are in Table 17. The interesting points are in bold. We see how the peak performance for M4 is at **x10y4**, almost 3x the default **x2y2**. Still, for our experiments we chose **x8y1** as it provided robust close-to-peak performance on all hosts we tested on. Note, **x8y1** requires a *wire format change*, so is not directly applicable to *stock* FSE-encoded data.

| **y \ x** | **2** | **4** | **6** | **8** | **10** | **12** | **16** |
|---|---|---|---|---|---|---|---|
| **y=1** | 828 | 1444 | 1762 | **2105** | 2001 | 2031 | 1641 |
| **y=2** | **854** | 1513 | 1734 | 1990 | 2014 | 1772 | 2284 |
| **y=4** | 857 | 1484 | 1706 | 2149 | **2308** | 1760 | 2060 |

Table 17: FSE wide-cursor decode throughput on M4 (MB/s), per cursor count $x$ and unroll $y$, at *p_maj=0.80*, 2880 B.

## G PivCo-Golomb

As discussed in Section 7.1, applying *pivoted coding* in other areas is an interesting research question. Coincidentally, just days before publishing this paper, a blog post appeared on fast Golomb decoding ([30]). Its author proposed a Tunstall-style approach, using a precomputed decoding table to quickly generate multiple output symbols (up to 8) from one byte of encoded data.

To investigate if we can apply our ideas in this field as well, we implemented **PivCo-Golomb**. It follows the approach very similar to PivCo-Huffman, except instead of a *tree*, it simply processes a *list* of bitmaps, starting from the bitmap for the last code bit all the way to the first bit. It uses a single primitive `merge_vec_cst_plus1` - which extends `merge_vec_cst` by adding `0x01` to produced values. The `cst` value used is `0xFF`. This way, when decoding a bitmap, values with `1` set will get an output symbol of `0x00`, and bitmap values of `0` will get the input symbols increased by 1.

| **avg code length** | **M4** (ns/symbol) | | | | **c8i** (ns/symbol) | | | |
|---|---|---|---|---|---|---|---|---|
| | naive | tunstall64 | t64-bf | PivCo-G | naive | tunstall64 | t64-bf | PivCo-G |
| 1.25 | 2.14 | 0.08 | 0.09 | 0.05 | 2.27 | 0.09 | 0.09 | 0.02 |
| 1.5 | 1.77 | 0.08 | 0.10 | 0.06 | 2.27 | 0.11 | 0.11 | 0.03 |
| 2 | 1.68 | 0.10 | 0.13 | 0.08 | 2.26 | 0.15 | 0.15 | 0.04 |
| 3 | 1.64 | 0.20 | 0.19 | 0.12 | 2.28 | 0.31 | 0.23 | 0.07 |
| 4 | 1.64 | 0.39 | 0.26 | 0.17 | 2.26 | 0.62 | 0.30 | 0.10 |
| 5 | 1.64 | 0.68 | 0.33 | 0.22 | 2.27 | 1.03 | 0.38 | 0.13 |

Table 18: Comparing Golomb-decoding implementation: naive, Tunstall-style 64-bit table from [30], branch-free version, and PivCo-Golomb

In Table 18 we can see that also in this application our approach can provide excellent performance. It is especially visible on longer average code lengths, where *tunstall64* (our name) from [30] slows down, mostly because of branch mispredictions. We also created a branch-free *t64-bf* version, which provides more stable performance. PivCo-Golomb's performance scales linearly with the length of the code, with the per-bit cost comparable to the `merge_vec_cst` performance we measured in Table 9.

This provides an interesting validation point that the *pivoted coding* idea could apply efficiently in other areas.